\documentclass[10pt,journal,compsoc]{IEEEtran}
\ifCLASSOPTIONcompsoc
  \usepackage[nocompress]{cite}
\else
  \usepackage{cite}
\fi
\ifCLASSINFOpdf
  \usepackage[pdftex]{graphicx}
  \DeclareGraphicsExtensions{.pdf,.jpeg,.png}
\else
\fi
\usepackage{amsmath}
\usepackage{amsfonts}
\usepackage{xspace}
\newcommand{\GDGD}{{$(SGD)^2$}\xspace}
\newcommand{\GDGDtext}{{Multicriteria Scalable Graph Drawing via Stochastic Gradient Descent}\xspace}
\usepackage[colorlinks=true,urlcolor=blue]{hyperref}
\usepackage{comment}

\usepackage{amsmath}
\usepackage{booktabs}
\usepackage{makecell}
\usepackage{tabularx}

\usepackage{todonotes}
\usepackage{xcolor,colortbl}

\usepackage{float}
\restylefloat{table}

\usepackage{cite}  

\usepackage{caption}
\usepackage{subcaption}
\usepackage[ruled,vlined]{algorithm2e}
\SetKwInput{KwInput}{Input}
\SetKwInput{KwOutput}{Output}
\SetKwProg{Fn}{Function}{:}{}

\newcommand{\blue}[1]{\textcolor{black}{#1}}

\begin{document}
%
\title{\GDGDtext, \GDGD}
%
%
%
%

\author{\IEEEauthorblockN{Reyan Ahmed, Felice De Luca, Sabin Devkota,  Stephen Kobourov, Mingwei Li}
\thanks{An earlier version of this paper appears in GD'20~\cite{ahmed2020gd}; in this extended version we use stochastic gradient descent which allows for multicriteria optimization on larger graphs.}
\IEEEauthorblockA{\\Department of Computer Science, University of Arizona, USA}}

\IEEEtitleabstractindextext{%
\begin{abstract}
Readability criteria, such as distance or neighborhood preservation, are often used to optimize node-link representations of graphs to enable the comprehension of the underlying data. 
With few exceptions, graph drawing algorithms typically optimize one such criterion, usually at the expense of others. 
We propose a layout approach, \GDGDtext, \GDGD, that can handle multiple readability criteria. 
%
\blue{
\GDGD can optimize any criterion that can be described by a differentiable function.
}
Our approach is flexible and can be used to optimize several criteria that have already been considered earlier (e.g., obtaining ideal edge lengths, stress, neighborhood preservation) as well as other criteria which have not yet been explicitly optimized in such fashion (e.g., node resolution, angular resolution, aspect ratio).
The approach is scalable and can handle large graphs. 
A variation of the underlying approach can also be used to optimize many desirable properties in planar graphs, while  maintaining planarity. 
%
\blue{
Finally, we provide quantitative and qualitative evidence of the effectiveness of \GDGD: we analyze the interactions between criteria, measure the quality of layouts generated from \GDGD as well as the runtime behavior, and analyze the impact of sample sizes.
The source code 
is available on github and we
also provide an interactive demo for small graphs.
}
\end{abstract}

}

\maketitle

\IEEEdisplaynontitleabstractindextext

%
\IEEEpeerreviewmaketitle

\IEEEraisesectionheading{
\section{Introduction}\label{sec:introduction}}
Graphs represent relationships between entities and visualization of this information is relevant in many domains. 
Several criteria have been proposed to evaluate the readability of graph drawings, including the number of edge crossings, distance preservation, and neighborhood preservation. 
Such criteria evaluate different aspects of the drawing and different layout algorithms optimize different criteria. 
It is challenging to optimize multiple readability criteria at once and there are few approaches that can support this.
Examples of approaches that can handle a small number of related criteria include the stress majorization framework of Wang et al.~\cite{wang2017revisiting}, which optimizes distance preservation via stress as well as ideal edge length preservation.
The Stress Plus X (SPX) framework of Devkota et al.~\cite{devkota2019stress}  can minimize the number of crossings, or maximize the minimum angle of edge crossings.
While these frameworks can handle a limited set of related criteria, it is not clear how to extend them to arbitrary optimization goals.
The reason for this limitation is that these frameworks are dependent on a particular mathematical formulation. 
For example, the SPX framework
was designed for crossing minimization, which can be easily modified to handle crossing angle maximization (by adding a cosine factor to the optimization function).
This ``trick" can be applied only to a limited set of criteria but not the majority of other criteria that are incompatible with the basic formulation. 

\begin{figure}[htbp]
  \includegraphics[trim={0 58 0 0}, clip, width=\columnwidth]{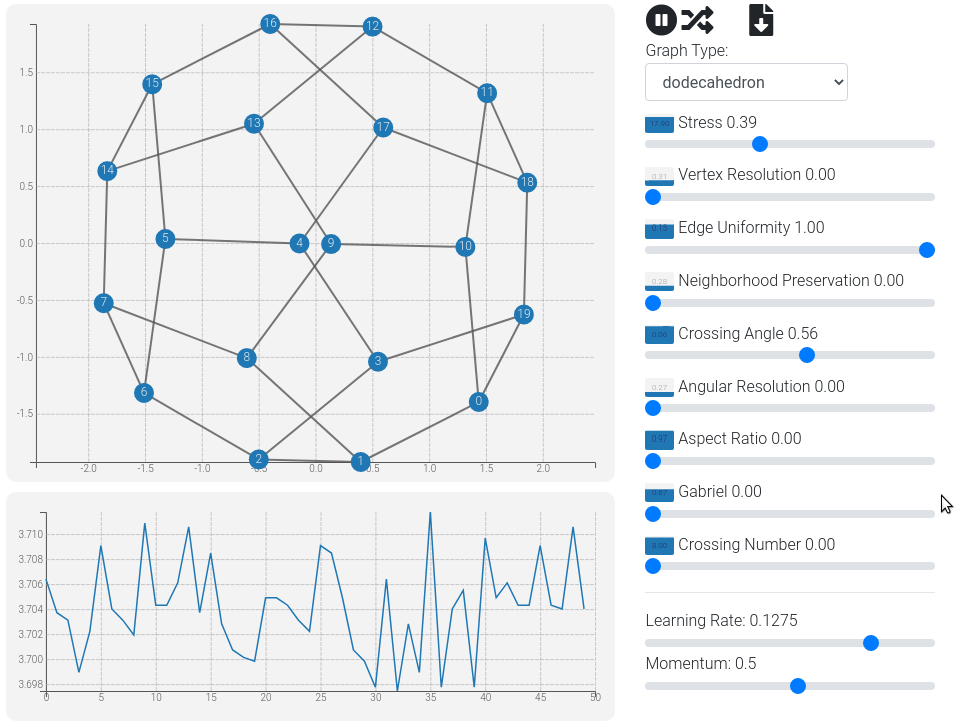}
\caption{\blue{An interactive prototype of \GDGD simutaneously optimizing stress, edge uniformity and crossing angles on a dodecahedron.}}
\label{fig:tensorflowjs-ui}
\end{figure}

In this paper, we propose a general approach, \GDGDtext, \GDGD, that can optimize a large set of drawing criteria, provided that the corresponding metrics that evaluate the criteria are differentiable functions. 
\blue{
If the criterion is not naturally differentiable, we design a differentiable surrogate function to approximate and optimize the original criterion.
In \GDGD, auto-differentiation tools are used for the gradient-based optimization.
}
To demonstrate the flexibility of the approach, we consider a set of nine criteria: minimizing stress, maximizing node resolution, obtaining ideal edge lengths, maximizing neighborhood preservation, maximizing crossing angle, optimizing total angular resolution, minimizing aspect ratio, optimizing the Gabriel graph property, and minimizing edge crossings. 
\blue{ 
We evaluate the effectiveness of our approach quantitatively and qualitatively  with evidence drawn from a set of experiments.
To illustrate the effectiveness and efficiency of multicriteria optimization, we evaluate the compatibility of every pair of criteria, measure the quality of each criterion, and demonstrate the distinctive looks of graph layouts under different drawing objectives.
We also evaluate the runtime performance and the impact of sample sizes used in the optimization, and compare our methods with existing ones.
}
We implemented our method with PyTorch. 
The code is available at: \url{https://github.com/tiga1231/graph-drawing/tree/sgd}.
For demonstration purposes, we also built an interactive prototype \blue{(that implements full-batch gradient descent on small graphs)} in JavaScript using tensorflow.js and D3.js, which is available on \url{http://hdc.cs.arizona.edu/~mwli/graph-drawing/}. 
This interactive prototype allows nodes to be moved manually and combinations of criteria can be optimized by selecting a weight for each; 
\blue{see Fig.~\ref{fig:tensorflowjs-ui}.}

\section{Related Work}
Many criteria associated with the readability of graph drawings have been proposed~\cite{ware2002cognitive}. Most graph layout algorithms are designed to (explicitly or implicitly) optimize a single criterion.
For instance, a classic  layout  criterion is stress minimization ~\cite{kamada_1989}, where stress is defined by $\sum\limits_{i < j}w_{ij} (|X_i-X_j| - d_{ij})^2$. Here, $X$ is a $n\times2$ matrix containing coordinates for the $n$ nodes, $d_{ij}$ is typically the graph-theoretical distance between two nodes $i$ and $j$ and $w_{ij}=d_{ij}^{-\alpha}$ is a normalization factor with $\alpha$ equal to $0, 1$ or $2$. Thus reducing the stress in a layout corresponds to computing node positions so that the actual distance between pairs of nodes is proportional to the graph theoretic distance between them. Optimizing stress can be accomplished by stress minimization, or stress majorization, which can speed up the computation~\cite{gansner2004graph}. In this paper we only consider drawing in the Euclidean plane, however, stress can be also optimized in other spaces such as the torus~\cite{chen2020doughnets}.

Stress minimization corresponds to optimizing the global structure of the layout, as the stress metric takes into account all pairwise distances in the graph. The t-SNET algorithm of Kruiger et al.~\cite{kruiger2017graph} directly optimizes neighborhood preservation, which captures the local structure of a graph, as the 
neighborhood preservation metric only  considers distances between pairs of nodes that are close to each other.
Optimizing local or global distance preservation can be seen as special cases of the more general dimensionality reduction approaches such as  multi-dimensional scaling~\cite{shepard1962analysis,kruskal1964multidimensional}. 

Purchase et al.~\cite{Purchase1997} showed that the readability of graphs increases if a layout has fewer edge crossings. The underlying optimization problem is NP-hard and several graph drawing contests have been organized with the objective of minimizing the number of crossings in the graph drawings~\cite{Abrego12,Buchheim13}. Recently several algorithms that directly minimize crossings have been proposed
~\cite{bennett2010,Radermacher18}. 

The negative impact on graph readability due to edge crossings can be mitigated if crossing pairs of edges have a large crossings angle~\cite{Argyriou2010,huang2014,huang2013,didimo2014crossangle}. Formally, the crossing angle of a straight-line drawing of a graph is the minimum angle between two crossing edges in the layout, and optimizing this property is also NP-hard.
Recent graph drawing contests have been organized with the objective of maximizing the crossings angle in graph drawings and this has led to several heuristics for this problem~\cite{Demel2018AGH,Bekos18}. 

The algorithms above are very effective at optimizing the specific readability criterion they are designed for, but they cannot be directly used to optimize additional criteria. This is a desirable goal, since optimizing one criterion often leads to poor layouts with respect to one or more other criteria: for example, algorithms that optimize the crossing angle tend to create drawings with high stress and no neighborhood preservation~\cite{devkota2019stress}.

Davidson and Harel~\cite{davidson1996drawing} used simulated annealing to optimize different graph readability criteria (keeping nodes away from other nodes and edges, uniform edge lengths, minimizing edge crossings). \blue{Huang et al.~\cite{huang2013} extended a force-directed algorithm to optimize crossing angle and angular resolution by incorporating two additional angle forces. The authors  show that in addition to  optimizing crossing angle and angular resolution, the algorithm also improves other desirable properties (average size of crossing angles, standard deviation of crossing angles, standard deviations of angular resolution, etc.). In a force-directed method similar to the algorithm proposed by Huang et al.~\cite{huang2013}, to optimize each criterion one needs to design a new force. The new force can be considered as a gradient update by hand, whereas \GDGD is a gradient descent based algorithm where the gradients are computed automatically using auto-differentiation tools.} 
Recently, several approaches have been proposed to simultaneously improve multiple layout criteria. Wang et al.~\cite{wang2017revisiting} propose a revised formulation of stress that can be used to specify ideal edge direction in addition to ideal edge lengths in a graph drawing. Wang et al.~\cite{wang2018structure} extended that stress formulation to produce structure-aware and smooth fish-eye views of graphs. 
Devkota et al.~\cite{devkota2019stress} 
also use a stress-based approach to minimize edge crossings and maximize crossing angles.
Eades et al.~\cite{10.1007/978-3-319-27261-0_41} provided a technique to draw large graphs while optimizing different geometric criteria, including the Gabriel graph property. Although the approaches above are designed to optimize multiple criteria, they cannot be naturally extended to handle other optimization goals. 
 

Constraint-based layout algorithms such as COLA~\cite{ipsepcola_2006, scalable_cola_2009}, can be used to enforce separation constraints on pairs of nodes to support properties such as customized node ordering or downward pointing edges. The coordinates of two nodes are related by inequalities in the form of $x_i \geq x_j + gap$ for a node pair $(i,j)$.  Dwyer et al.~\cite{dwyer2009constrained} use gradient projection to handle these constraints, be moving nodes as little as needed to satisfy the inequalities/ equalities after each iteration of the layout method. The gradient projection method has been extended to also handle  non-linear constraints~\cite{dwyer2009layout}.  These hard constraints are powerful but a bit restrictive and are different from \blue{the soft constraints} in our \GDGD framework. 

\section{The \GDGD Framework}
\begin{figure}[t]
\centering
  \includegraphics[width=\linewidth]{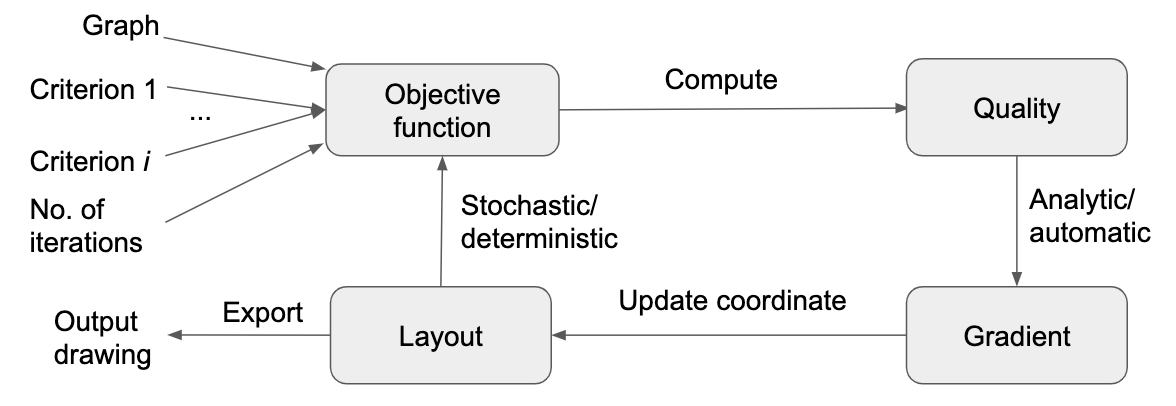}
  \caption{The \GDGD framework: Given a graph and a set of criteria (with weights), formulate an objective function based on the selected set of criteria and weights. Then compute the quality (value) of the objective function of the current layout of the graph. Next, generate the gradient (analytically or automatically). Using the gradient information, update the coordinates of the layout. Finally, update the objective function based on the layout via regular or stochastic gradient descent. This process is repeated for a fixed number of iterations.}
  \label{fig:gdgdframework}
\end{figure}

%
%
The \GDGD framework is a general optimization approach to generate a layout with any desired set of aesthetic metrics, provided that they can be expressed by a smooth function. The basic principles underlying this framework are simple. 
The first step is to select a set of layout readability criteria and loss functions that measure each of them. 
Then we define the function to optimize as a linear combination of the loss functions for each individual criterion.
Finally, we iterate the gradient descent steps, from which we obtain a slightly better drawing at each iteration.
Fig.~\ref{fig:gdgdframework} depicts the framework of \GDGD: 
Given any graph with $n$ nodes and a readability criterion $Q$, we design a loss function $L_{Q}: \mathbb{R}^{n \times 2} \to \mathbb{R}$ that maps the current layout $X \in \mathbb{R}^{n \times 2}$ to a measure $L_{Q}(X)$ with respect to the readability criterion. 
Then we combine multiple loss functions from different criteria into a single one by taking a weighted sum, $L(X) = \Sigma_{Q}w_Q L_{Q}(X)$, where a lower value is always desirable.
At each iteration, a slightly better layout can be found by taking a small ($\epsilon$) step along the (negative) gradient direction: $X^{(new)} = X - \epsilon \cdot \nabla\; L(X)$.
\blue{Algorithm~\ref{alg:gd2} summarises the \GDGD optimization procedure.}

\begin{algorithm}[t]
\DontPrintSemicolon
\caption{\blue{The \GDGD Algorithm}\label{alg:gd2}}
\KwInput{
    \\
    $G = (V,E)$ \tcp{graph}
    $C = \{\dots c \dots\}$ \tcp{criteria}
    $S: c \mapsto s_c$ \tcp{sample sizes for each $c$}
    $L_c: \mathbb{R}^{s_c \times 2} \to \mathbb{R}_+$ \tcp{loss functions}
    $maxiter \in \mathbb{Z}_+$ \tcp{number of iterations}
    $W: c \mapsto w_c$, where $w_c: [1, maxiter] \to \mathbb{R_+}$ \tcp{weight schedules}
    $\eta: [1, maxiter] \to \mathbb{R_+}$ \tcp{learning rate}
    $q$ \tcp{criterion for safe update}
    $Q_{q}$ \tcp{quality measure of $q$}
}
\KwOutput{
    $X$ \tcp{layout that optimizes multiple criteria}
}
\Fn{Layout($G; C, S, W, maxiter, \eta$)}{
    $X \leftarrow$ InitializeLayout($G$)\;
    \If{`crossings' $\in C$}{
        $cd \leftarrow$ InitializeCrossingDetector()\;
    }
    \For {$t = 1, \dots, maxiter$}{
        $l \leftarrow 0$\;
        \For {$c \in C$ s.t. $w_c(t) > 0$}{
            $sample \leftarrow$ Sample($c, s_c$)\;
            \If{c == `crossings'}{
                UpdateCrossingDetector($cd, sample$)\;
                $l_{c} \leftarrow L_{c}(sample; G, cd)$\;
            }
            \Else{
                $l_{c} \leftarrow L_{c}(sample; G)$\;
            }
            $l \leftarrow l + w_c(t) \cdot l_c$\;
        }
        
        \If{`Safe update' is enabled}{
            $X_{prev} \leftarrow X$\;
            $X_{new} \leftarrow X - \eta(t) \cdot \nabla_{X} l$\;
            $X \leftarrow$ SafeUpdate($X_{prev}, X_{new}; G, Q_q$) \tcp{Alg.  \ref{alg:safe_update_2}}
        }
        \Else{
            $X \leftarrow X - \eta(t) \cdot \nabla_{X} l$\;
        }
    }
    Return X\;
}
\end{algorithm}

\subsection{Gradient Descent Optimization}

There are different kinds of gradient descent algorithms. 
The standard method considers all nodes, computes the gradient of the objective function, and updates node coordinates based on the gradient. 
Some objectives may consider all the nodes in every step. 
For example, the basic stress formulation~\cite{kamada_1989} falls in this category.
To compute the gradient for optimization, one has to iterate through all the nodes which makes it not scalable to very large graphs.
Fortunately,most of these objectives can be decomposed into optimization over only subsets of nodes. 
Consider stress minimization again, if we sample a set of node pairs randomly and minimize the stress between the nodes in each pair, the stress of the whole graph is also minimized~\cite{zheng2018graph}.
\blue{This approach is known as stochastic gradient descent (SGD) and we use this idea extensively.}
In section~\ref{sect:properties-and-measures}, we specify the objective loss functions and sampling methods we used for each readability criterion we consider. 

Not all readability criteria come naturally in the form of differentiable functions. 
We cannot compute the gradient of or apply SGD on non-differentiable functions.
In cases that the original objective is continuous but not everywhere differentiable e.g., a 'hinge' function f(x)=max(0,x), we can compute the subgradient and update the objective based on the subgradient.
Hence, as long as the function is continuously defined on a connected component in the domain, we can apply the subgradient descent algorithm.

When a function is not defined in a connected domain, we can introduce surrogate loss functions to `connect the pieces'.
For example, when optimizing neighborhood preservation we maximize the Jaccard similarity between graph neighbors and nearest neighbors in graph layout.
However, Jaccard similarity is only defined between two binary vectors.
To solve this problem we extend Jaccard similarity to all real vectors by its Lov\'{a}sz extension~\cite{berman2018lovasz} and apply that to optimize neighborhood preservation. 
%
An essential part of gradient descent based algorithms is to compute the gradient/subgradient of the objective function. 
In practice, it is not necessary to write down the gradient analytically as it can be computed automatically via (reverse-mode) automatic differentiation~\cite{griewank2008evaluating}. 
Deep learning packages such as Tensorflow~\cite{abadi2016tensorflow} and PyTorch~\cite{paszke2019pytorch} apply automatic differentiation to compute the gradient of complicated functions. 
%
%
%

\blue{
Most of the objective functions that we consider here are not convex and do not have unique global minimizers. 
Therefore, even though SGD is known to converge (to at least a local optimum) in relatively relaxed settings~\cite{gower2019sgd, bassily2018exponential}, few optimization objectives are guaranteed to find the global optimum. 
In particular, unlike methods such as stress majorization~\cite{gansner2004graph}, most of our optimization objectives are not guaranteed to converge to the global optimum.
Meanwhile, most of the objective functions for which SGD works well in practice (e.g., in deep learning) are neither convex nor have unique global minimizers~\cite{li2017visualizing}. 
With this in mind, we follow the common practice of applying an annealing process, if necessary, to ensure  convergence (to a possibly local minimum).
}

When optimizing multiple criteria simultaneously, we combine them via a weighted sum. 
However, choosing a proper weight for each criterion can be tricky. 
Consider, for example, maximizing crossing angles and minimizing stress simultaneously with a fixed pair of weights. 
At the very early stage, the initial drawing may have many crossings and stress minimization often removes most of the early crossings. 
As a result, maximizing crossing angles in the early stage can be harmful as it moves nodes in directions that contradict those that come from stress minimization.
Therefore, a well-tailored \textit{weight scheduling} is needed for a successful outcome.
Continuing with the same example, a better outcome can be achieved by first optimizing stress until it converges, and later adding weights for the crossing angle maximization. 
To explore different ways of scheduling, we provide an interface that allows manual tuning of the weights. 
\blue{
We consider weight schedules for different criteria sets in Section~\ref{sect:analysis-of-qualities}.
}

\subsection{Implementation}
We implemented the \GDGD framework in Python.
In particular we used PyTorch~\cite{paszke2019pytorch} automatic differentiation, NetworkX~\cite{networkx} for processing graphs, and matplotlib~\cite{matplotlib} for drawing.
The code is available at \url{https://github.com/tiga1231/graph-drawing/tree/sgd}.
To demonstrate our method with small graphs, we have provided an interactive tool written in JavaScript \footnote{ \url{http://hdc.cs.arizona.edu/~mwli/graph-drawing/}}, where we used the automatic differentiation tools in tensorflow.js~\cite{mlsys2019_154} and the drawing library D3.js~\cite{2011-d3}.

\section{Properties and Measures}\label{sect:properties-and-measures}
In this section we specify the aesthetic goals, definitions, quality measures and loss functions for each of the $9$ graph drawing properties we optimized: stress, ideal edge lengths, neighborhood preservation, crossing number, crossing angle, aspect ratio, angular resolution, node resolution and Gabriel graph property.
Other standard graph notation is summarized in Table~\ref{table:notations}.
In each subsection, we first define our loss function for the entire graph. 
For small graphs, one can apply (full-batch) gradient descent directly on this loss.
To speed up our method for larger graphs, we sample portions of our loss functions at each iteration and apply (mini-batch) stochastic gradient descent on them. 
\blue{
The definition of a sample can be different for each criterion. 
For example, for stress minimization we sample pairs of nodes; for ideal edge length, we sample edges.
Hence, the sample sizes of different criteria can be set independently.
Moreover, when the sample size for a certain criterion exceeds the number of possible samples, our method is automatically equivalent to (full-batch) gradient descent for that criterion.
In Section~\ref{sect:analysis-of-sample-size}, we discuss the effect of the sample sizes on the convergence rates. 
The analysis has helped us set the default values for each readability criterion.
In general, for each criterion we sample mini-batches from a pool of all sample points (e.g., all pairs of nodes for stress, all edges for ideal edge length) without replacement, and `refill the pool' when all sample points are drawn.
In practice, we shuffle the list of data points, draw mini-batches from the list in consecutive order, and re-shuffle the list once every data point is drawn.
}
\begin{table}[h]
\resizebox{\columnwidth}{!}{%
    \begin{tabular}{l|l}
    \toprule
    Notation & Description\\
    \midrule
    $G$ & Graph\\
    $V$ & The set of nodes in $G$, indexed by $i$, $j$ or $k$\\
    $E$ & The set of edges in $G$, indexed by a pair of nodes $i,j$ in $V$\\
    $n=|V|$ & Number of nodes in $G$\\
    $m$ & sample size for a certain criterion in SGD\\
    $|E|$ & Number of edges in $G$\\
    $Adj$ and $A_{i,j}$ & Adjacency matrix of $G$ and its $(i,j)$-th entry\\
    $d_{ij}$ & Graph-theoretic distance between node $i$ and $j$\\
    $X_{n \times 2}$ & 2D-coordinates of nodes in the drawing\\
    $||X_i - X_j||$ & The Euclidean distance between nodes $i$ and $j$ \\
    $\theta_i$ & $i^{th}$ crossing angle\\
    $\varphi_{ijk}$ & Angle between incident edges $(i,j)$ and $(j,k)$\\
    \bottomrule
    \end{tabular}
    \caption{Graph notation used in this paper.}
    \label{table:notations}
}
\end{table}

\subsection{Stress}
We minimize stress, $L_{ST}$, to draw a graph that matches the Euclidean distances between pairs of nodes in the drawing to their graph theoretic distances. 
Following the original definition of stress~\cite{kamada_1989}, we minimize
\begin{align}
L_{ST} = \sum\limits_{i<j}\;w_{ij}(||X_i - X_j||_2 - d_{ij})^2 \label{eq:loss-stress}
\end{align}
Where $d_{ij}$ is the graph-theoretical distance between nodes $i$ and $j$, $X_i$ and $X_j$ are the coordinates of nodes $i$ and $j$ in the layout. 
The normalization factor $w_{ij}=d_{ij}^{-2}$ balances the influence of short and long distances: the longer the graph theoretic distance, the more tolerance we give to the discrepancy between two distances.
When comparing two drawings of the same graph with respect to stress, a smaller value (lower bounded by $0$) corresponds to a better drawing. 
To work with large graphs, we take the mean loss for any pairs of nodes turn it into the expectation of stress
\begin{align}
\hat{L}_{ST} = \mathbb{E}_{i\neq j}\; [w_{ij}(||X_i - X_j||_2 - d_{ij})^2] \label{eq:loss-stress-expectation}
\end{align}
\blue{The quality measure for stress, $Q_{ST}$, is equal to the loss $\hat{L}_{ST}$ over all pairs of nodes.}
In each SGD iteration we minimize the loss by sampling a number of node pairs. 
\blue{
Since the expectation of the gradient of the sample loss equals the true loss, we can use the gradient of the sample loss as an estimate of the true gradient and update the drawing through SGD accordingly.
In each SGD iteration, we sample $m$ pairs of nodes. 
By default, we set $m=32$ based on our experiments with different sample sizes in Section~\ref{sect:analysis-of-sample-size}.
Before a round that goes over all 
pairs of nodes, we shuffle a list of node-pairs and take mini-batches from the shuffled list. 
This guarantees that we process every pair of nodes exactly once per round.
}

\subsection{Ideal Edge Length}
Given a set of ideal edge lengths $\{l_{ij}: (i,j) \in E\}$ we minimize the variance from the ideal lengths:

\begin{align}
L_{IL} &= \sum\limits_{(i,j) \in E}\;  
(\frac{||X_i - X_j|| - l_{ij}}{l_{ij}})^2 \label{eq:loss-ideal-edge-length}
\end{align}
For unweighted graphs, by default we use $1$ as the ideal edge length for all edges $(i,j) \in E$. 

As with stress minimization, for large graphs we replace the summation by the expectation and estimate it through sampling the edges. 
 \begin{align}
 \hat{L}_{IL} &= \mathbb{E}_{(i,j) \in E}[
     (\frac{||X_i - X_j|| - l_{ij}}{l_{ij}})^2
 ]
 \label{eq:loss-ideal-edge-length-expectation}
 \end{align}
\blue{
The quality measure $Q_{IL} = \hat{L}_{IL}$ is lower bounded by $0$ and a lower score yields a better layout.
Similar to the sampling strategy for stress, here we keep a list of all edges in random order, draw mini-batches (by default, of size $m=32$) from it, and re-shuffle the list after all edges are processed once.}

\subsection{Neighborhood Preservation}
\label{sec:neighbor}
Neighborhood preservation aims to keep adjacent nodes close to each other in the layout.
Similar to Kruiger et al.~\cite{kruiger2017graph}, the idea is to have the $k$-nearest (Euclidean) neighbors (k-NN) of node $i$ in the drawing to align with the $k$ nearest nodes (in terms of graph distance from $i$). 
Here we choose $k$ to be the degree of node - for nodes of different degrees, we consider a different number of neighbors.
A natural quality measure for the alignment is the Jaccard index between the two pieces of information. Let, $Q_{NP} = JaccardIndex(K, Adj) = \frac{|\{(i,j): K_{ij}=1 \text{ and } A_{ij}=1\}|}{|\{(i,j): K_{ij}=1 \text{ or } A_{ij}=1\}|}$, where $Adj$ denotes the adjacency matrix and the $i$-th row in $K$ denotes the $k$-nearest neighborhood information of $i$:
$K_{ij} = 1$ if $j$ is one of the k-nearest neighbors of $i$ and $K_{ij}$ = 0 otherwise.

To express the Jaccard index as a differentiable minimization problem, 
first, we express the neighborhood information in the drawing as a smooth function of node positions $X_i$ and store it in a matrix $\hat{K}$.
In $\hat{K}$, a positive entry $\hat{K}_{i,j}$ means node $j$ is one of the k-nearest neighbors of $i$, otherwise the entry is negative.
Next, we take a differentiable surrogate function of the Jaccard index, the Lov\'{a}sz hinge loss (LHL) given by Berman et al.~\cite{berman2018lovasz}, to make the Jaccard loss optimizable via gradient descent.
We minimize

\begin{align}
L_{NP} &= LHL(\hat{K}, Adj)\label{eq:lovasz-hinge}
\end{align}
where $\hat{K}$ denotes the $k$-nearest neighbor estimation. 
For simplicity, let $d_{i,j}=||X_i - X_j||$ denote the Euclidean distance between node $i$ and $j$, then we design $\hat{K}$ as:

\begin{align}
\hat{K}_{i,j} &= 
\left\{\begin{array}{ll}
  -(d_{i,j} - \frac{d_{i,\pi_k} + d_{i,\pi_{k+1}}}{2} ) & \text{ if } i \neq j\\
  0 & \text{ if } i=j\\
\end{array}\right.\label{eq:neighbor-pred}
\end{align}
where $\pi_{k}$ denotes the $k^{th}$ nearest neighbor of node $i$.
In other words, for every node $i$, we treat the average distance to its $k^{th} $ and $(k+1)^{th}$ nearest neighbor as a threshold, and use it to measure whether node $j$ is in the neighbor or not. 
Note that $d_{i,j}$, $d_{i,\pi_k}$ and $d_{i,\pi_{k+1}}$ are all smooth functions of node positions in the layout, so $\hat{K}_{i,j}$ is also a smooth function of node positions $X$. 
Furthermore, $\hat{K}_{i,j}$ is positive if node $j$ is a k-NN of node $i$, otherwise it is negative, as is required by LHL~\cite{berman2018lovasz}.

In order to handle large graphs we sample nodes for stochastic gradient descent. 
However, note that the nearest neighbors $\pi_k$ and $\pi_{k+1}$ in $\hat{K}_{i,j}$ depend on distances from all nodes.
To derive a reliable estimation of the Jaccard index, instead of letting $k$ equal to the degree of node $i$ in the full graph, we need $k$ equal to the degree of the subgraph that we sample.
In other words, in every gradient descent iteration we sample a subgraph from the full graph and compute $LHL$ of the subgraph.
In practice, we randomly select a small set of $m$ nodes \blue{(by default, $m=16$)}, along with nodes that are $1$ or $2$ hops away from any of them.
We also include a fraction of nodes that are not already in the sample.
We extract the subgraph induced by this set of nodes and apply stochastic gradient descent.

%
%
\subsection{Crossing Number}
Reducing the number of edge crossings is one of the classic optimization goals in graph drawing, known to affect readability~\cite{Purchase1997}.
\blue{
Shabbeer et al.~\cite{bennett2010}, employed an expectation-maximization-like algorithm to minimize the number of crossings. 
Since two edges do not cross if and only if there exists a line that separates their extreme points, they trained many support vector machine (SVM) classifiers to separate crossing pairs and use the classifiers as a guide to eliminate crossings.
Since one has to train as many SVM classifiers as the number of crossings in the graph and knowledge learned by one SVM does not naturally transfer to another, we found that this approach does not work well on large graphs.
With this in mind we modified our initial approach to that of Tiezzi et al.\cite{tiezzi2021graph}, which uses Graph Neural Networks to reduce the number of crossings in two steps.
First, they train a generic neural network to predict if any two edges cross each other.
Since neural networks are differentiable, the well-trained edge crossing predictor from this step will serve as a guide to gradient descent steps later on.
In the second step they train a Graph Neural Network and use the edge crossing predictor as a guide to improve the layout.
Our method uses only the first step above and utilizes a different  training strategy.
Instead of training the edge crossing predictor using a synthetic dataset before the layout optimization, we train the crossing predictor directly on the current graph layout while simultaneously updating the node coordinates in the same graph, using the crossing predictor as a guide.
}
\blue{
Formally, let $f_\beta$ denote a neural network with trainable parameters $\beta$ that takes the coordinates of the four nodes of any two edges $X^{(i)} \in \mathbb{R}^{4 \times 2}$ and outputs a scalar from the $(0,1)$ interval.
An output close to $0$ means ``no crossing'' and one close to $1$ means ``crossing''.
In practice, $f_{\beta}$ is a simple multi-layer perceptron (MLP) with batch normalization~\cite{ioffe2015batch} and LeakyReLU activation.
To train a neural crossing detector $f_{\beta}$, we feed different edge pairs $X^{(i)} \in \mathbb{R}^{4 \times 2}$ to approximate the ground truth $t^{(i)} \in \{0, 1\}$ where $0$ means ``no crossing'' and $1$ means ``crossing''.
We optimize the parameters $\beta$ to minimize the cross entropy (CE) loss $L_{\beta}$ between the prediction $f_{\beta}(X^{(i)})$ and the ground truth $t^{(i)}$, averaging over a sample of $n$ instances of edge pairs:
$$
L_{\beta}(\beta;X^{(1)} \dots X^{(n)}) = \frac{1}{n}\sum\limits_{i=1}^n 
CE(f_\beta (X^{(i)}), t^{(i)})
$$
where
\begin{align}
CE(y, t) := - t \cdot log(y) - (1-t) \cdot log(1-y)  \label{eq:cross-entropy}
\end{align}
We use the neural crossing detector $f_\beta$ to construct a differentiable surrogate loss function for crossing minimization.
Specifically, given a well-trained $f_\beta$, we can reduce the number of crossings in a layout by minimizing the cross entropy between the prediction of edge pairs $f_\beta(X^{(i)})$ and the desired target (i.e., no crossing $t=0$):
\begin{align}
L_{CR}(X; \beta) = \frac{1}{n}\sum\limits_{i=1}^n CE(f_\beta (X^{(i)}), 0)
\end{align}
In practice, we minimize $L_{\beta}$ and $L_{CR}$ simultaneously in each \GDGD iteration.
We first improve the neural crossing predictor using a sample of edge pairs from the graph.
For simplicity, we describe the training by SGD, although in practice one can utilize any SGD variants (e.g. SGD with momentum, ADAM\cite{kingma2014adam} or RMSProp~\cite{Tieleman2012}) to train the predictor more efficiently.
\begin{align}
\beta^{(new)} = \beta - \epsilon' \cdot \nabla L_{\beta}
\end{align}
In the meantime we update the layout in a similar manner:
\begin{align}
X^{(new)} = X - \epsilon \cdot \nabla L_{CR}
\end{align}
Although one could improve the neural crossing predictor by multiple steps in every \GDGD iteration, we found little difference when varying the number of steps.
Therefore, we only take one step to improve the neural crossing predictor in every \GDGD iteration.
As with other criteria, we randomly draw mini-batches (by default, of size $m=128$) and iterate through all edge pairs over the course of the SGD iterations. 
}

\blue{
When a graph layout does not have many crossings (e.g., a stress-minimized layout of a near-planar graph), this sampling strategy is not efficient.
In that case, we use an efficient algorithm (Bentley-Ottmann~\cite{bentley1979algorithms}) to find all crossing edges in a graph, and sample a mini-batch of crossings.
Since finding all crossings can be slow for large graphs, we only do this once every few iterations and reuse the finding across a few iterations.
Specifically, when we draw mini-batches from the pool of all crossings, we recompute all crossings again once the pool is drained.
To evaluate the quality we simply count the number of crossings.
}

\subsection{Crossing Angle Maximization}
When edge crossings are unavoidable, the graph drawing can still be easier to read when edges cross at angles close to 90 degrees~\cite{ware2002cognitive}. 
Heuristics such as those by Demel et al.~\cite{Demel2018AGH} and Bekos et al.~\cite{Bekos18} have been proposed and have been successful in graph drawing challenges~\cite{devanny2017graph}.
We use an approach similar to the force-directed algorithm given by Eades et al.~\cite{eades2010force} and minimize the squared cosine of crossing angles: 
\begin{align}
L_{CAM} 
= \sum_{\substack{\text{all crossed edge pairs }\\(i,j), (k,l) \in E}} 
(\frac{\langle X_{i}-X_{j}, X_{k}-X_{l}\rangle}{|X_{i}-X_{j}|\cdot|X_{k}-X_{l}|})^2
\end{align}
We evaluate quality by measuring the worst (normalized) absolute discrepancy between each crossing angle $\theta$ and the target crossing angle (i.e. 90 degrees):
$
Q_{CAM} = \max_{\theta} |\theta - \frac{\pi}{2}| / \frac{\pi}{2}
$.
As with crossing numbers, for large graphs we sample a subset \blue{(by default, of size $m=16$)} of edge pairs and consider their crossing angles if the edge pair cross each other. 
Again, if there are not many crossing pairs we use an efficient algorithm to find all crossings.
When optimizing the number of crossings and crossing angles simultaneously, we sample from the same pool of crossings formed via the Bentley-Ottmann algorithm. 

\subsection{Aspect Ratio}

Good use of drawing area is often measured by the aspect ratio~\cite{duncan1998balanced} of the bounding box of the drawing, with 1:1 as the optimum.
\blue{
The idea here is to consider different rotations of the current layout and try to ``squarify" the corresponding bounding boxes.
In practice, 
we rely on the singular values of the matrix of node coordinates to approximate the worst aspect ratio.
Formally, assume vertex coordinates are centered with zero mean and let $X$ denote the collection of (centered) vertex coordinates as rows in a matrix.
Since the coordinates are two dimensional, $X$ has only two (non-zero) singular values, denoted by $\sigma_1$ and $\sigma_2$ and each measures the standard deviation of the layout along with two orthogonal directions.
Then we approximate the aspect ratio using the quotient of the two singular values of $X$ and encourage the ratio to be close to the target ratio $r=1$ using the cross entropy (CE) in Eq.~\ref{eq:cross-entropy}:
$$
L_{AR} = CE(\frac{\sigma_2}{\sigma_1}, r)
$$
Note that although we only consider 1:1 ratios, the formulation of cross entropy let us consider arbitrary ratios.
During mini-batch SGD, we simply sample a subset of nodes (by default, of size $m=128$) and use the singular values of the matrix formed by the subset to optimize the aspect ratio.  
}

Finally, we evaluate the drawing quality by measuring the worst aspect ratio on a finite set of rotations. 
The quality score ranges from 0 to 1. 
In our case, 1 is optimal and the minimal ratio among different rotations is the worst:
$
Q_{AR} = \min_{
\theta \in \{
    \frac{2\pi k}{N}, \text{ for } k=0, \cdots (N-1)
  \}
} \frac{\min(w_{\theta}, h_{\theta})}{\max(w_{\theta}, h_{\theta})}
$, 
where $N$ is the number of rotations sampled (e.g., $N=7$), and $w_{\theta}$, $h_{\theta}$ are the width and height of the bounding box when rotating the drawing around its center by an angle $\theta$.

\subsection{Angular Resolution}
Distributing edges adjacent to a node makes it easier to perceive the information presented in a node-link diagram~\cite{huang2013}. 
Angular resolution~\cite{Argyriou2010}, defined as the minimum angle between incident edges, is one way to quantify this goal. 
Formally, 
$
ANR = \min_{j \in V} \min_{(i,j),(j,k) \in E} \varphi_{ijk}
$,
where $\varphi_{ijk}$ is the angle formed by between edges $(i,j)$ and $(j,k)$.
Note that for any given graph, an upper bound of this quantity is $\frac{2\pi}{d_{max}}$ where $d_{max}$ is the maximum degree of nodes in the graph.
Therefore in the evaluation, we will use this upper bound to normalize our quality measure to $[0,1]$, i.e. 
$
Q_{ANR} = \frac{ANR}{2\pi / d_{max}}
$.
To achieve a better drawing quality via gradient descent, we define the angular energy of an angle $\varphi$ to be $e^{-s \cdot \varphi}$, where $s$ is a constant controlling the sensitivity of angular energy with respect to the angle (by default $s=1$), and minimize the total angular energy over all incident edges:

\begin{align}
L_{ANR} = \sum_{(i,j),(j,k) \in E} e^{-s \cdot \varphi_{ijk}} 
\end{align}

When the graph is large, it is expensive to compute the energy of all pairs of incident edges. 
Therefore, in \GDGD we randomly sample a minibatch of pairs of incident edges \blue{(by default, of size $m=128$)} and minimize the energy of the sample accordingly.

\begin{figure}[t]
    \centering
    \includegraphics[width=0.50\columnwidth]{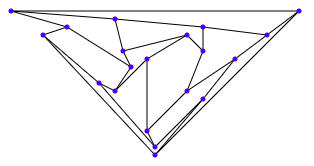}
    \includegraphics[width=0.35\columnwidth]{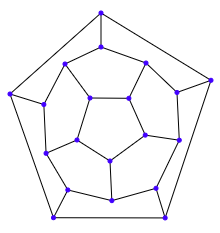}
    \caption{Optimizing Planar Graphs: (a) An initial layout without crossings, (b) A layout after optimizing stress while maintaining planarity.}
    \label{fig:maintian_planarity}



\end{figure}

\begin{figure}[t]
  \includegraphics[trim={5 10 25 10}, clip, width=0.49\columnwidth]{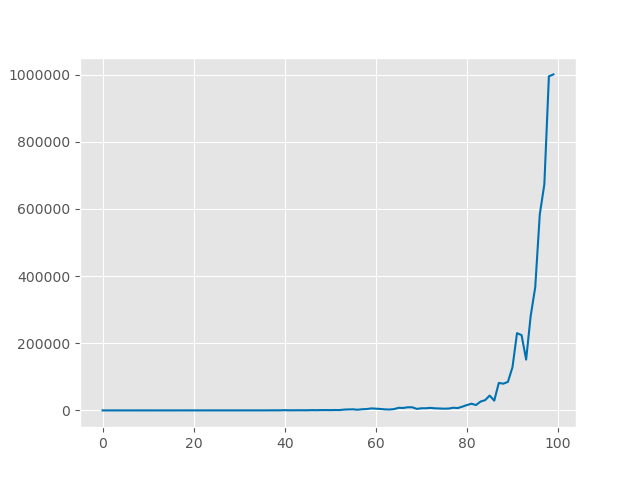}
  \includegraphics[trim={5 10 25 10}, clip, width=0.49\columnwidth]{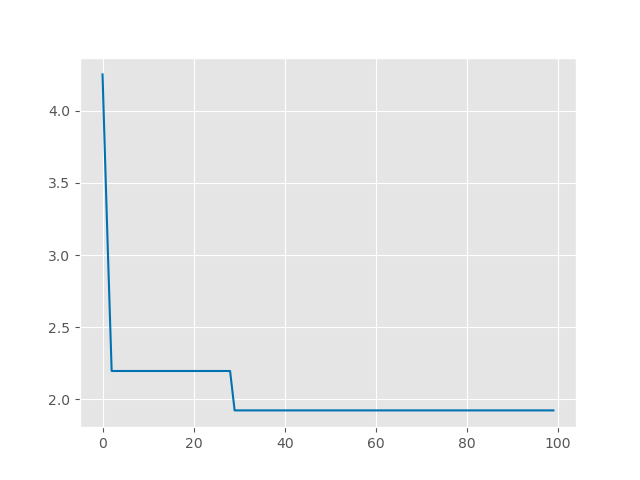}
\caption{(a) The edge uniformity loss is increasing when we optimize the stress of a nested triangular graph, (b) The loss is decreasing when we update the coordinates carefully.}
\label{fig:maintian_EU_triangular}
\end{figure}

\subsection{Node Resolution}
Good node resolution is associated with the ability to  distinguish different nodes by preventing nodes from occluding each other. 
Node resolution is typically defined as the minimum Euclidean distance between two nodes in the drawing~\cite{chrobak1996convex,schulz2011drawing}. 
However, in order to align with the units in other objectives such as stress, we normalize the minimum Euclidean distance with respect to a reference value.
Hence we define the node resolution to be the ratio between the shortest and longest  distances between pairs of nodes in the drawing, 
$
VR = \frac{\min_{i \neq j}||X_i - X_j||}{r \cdot d_{max}}
$,
where $d_{max} = \max_{k,l}||X_k - X_l||$. 
To achieve a certain target resolution $r \in [0,1]$ by minimizing a loss function, we minimize 

\begin{align}
L_{VR} = \sum_{i,j \in V, i \neq j}
max(
    0, 
    ( 1 - \frac{||X_i - X_j||}{r \cdot d_{max}})^2
)
\end{align}
In practice, we set the target resolution to be $r=\frac{1}{\sqrt{|V|}}$, where $|V|$ is the number of nodes in the graph. 
In this way, an optimal drawing will distribute nodes uniformly in the drawing area.
Each term in the summation vanishes when the distance between two nodes meets the required resolution $r$, otherwise it is greater than zero.
In the evaluation, we report, as a quality measure, the ratio between the actual and target resolution and cap its value between $0$ (worst) and $1$ (best).
$
Q_{VR} = \min(1.0, \frac{\min_{i,j} ||X_i - X_j||}{r \cdot d_{max}})
$

For large graphs, we sample a subset of nodes \blue{(by default, of size $m=256$)} and compute the approximate loss of node resolution on the sample.

\subsection{Gabriel Graph Property}
A graph is a Gabriel graph if it can be drawn in such a way that any disk formed by using an edge in the graph as its diameter contains no other nodes. 
Not all graphs are Gabriel graphs, but drawing a graph so that as many of these edge-based disks are empty of other nodes has been associated with good readability~\cite{10.1007/978-3-319-27261-0_41}.
This property can be enforced by a repulsive force around the midpoints of edges.
Formally, we establish a repulsive field with radius 
$r_{ij}$ equal to half of the edge length, around the midpoint $c_{ij}$ of each edge $(i,j) \in E$, and we minimize the total potential energy:

\begin{align}
L_{GA} = \sum_{
\substack{
  (i,j) \in E,\\
  k \in V \setminus \{i,j\}
}}
max(0, r_{ij} - |X_k - c_{ij}|) \; ^ 2
\label{eq:gabriel}
\end{align}
where 
$
c_{ij} = \frac{X_i + X_j}{2}
$ and
$
r_{ij} = \frac{|X_i - X_j|}{2}
$.
We use the (normalized) minimum distance from nodes to centers to characterize the quality of a drawing with respect to Gabriel graph property:
$
Q_{GA} = \min (1, \min_{(i,j) \in E, k \in V}\frac{|X_k - c_{ij}|}{r_{ij}})
$.

For large graphs, we sample a mini-batch of node-edge pairs \blue{(by default, of size $m=64$)} and compute the approximate loss from the sample.

\begin{algorithm}[t]
\DontPrintSemicolon
\caption{\blue{Update coordinates without quality decline}}\label{alg:safe_update_2}
\Fn{SafeUpdate($X_{prev}, X_{new}; G, Q_q$)}{
    $X \leftarrow X_{prev}$\;
    $q_0 \leftarrow Q_q(X; G)$\;
    \For{each node $u \in V$}{
        $X[u] \leftarrow X_{new}[u]$\;
        $q_{tmp} \leftarrow Q_q(X)$\;
        \If{QualityDeclines($q_0, q_{tmp}$)}{
            $X[u] \leftarrow X_{prev}[u]$\;
        }
    }
    return X\;
}
\end{algorithm}

\begin{figure*}[thbp]
\centering
\begin{subfigure}[b]{0.32\linewidth}
    \centering
    \includegraphics[width=.49\linewidth]{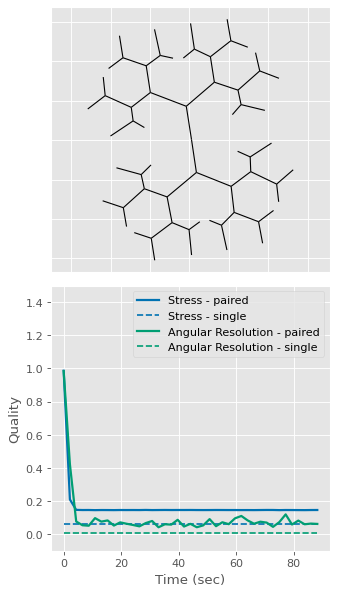}
    \includegraphics[width=.49\linewidth]{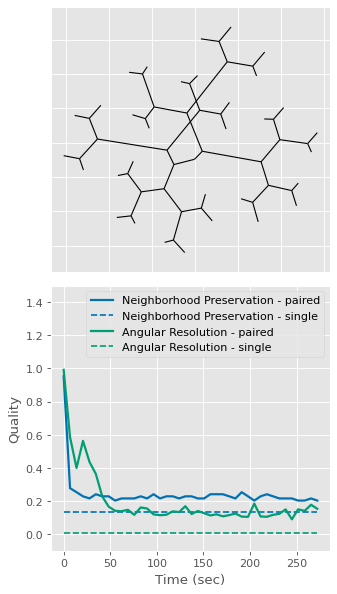}
    \caption{\blue{Compatible pairs}}
\end{subfigure}
\hfill
\begin{subfigure}[b]{0.32\linewidth}
    \centering
    \includegraphics[width=.49\linewidth]{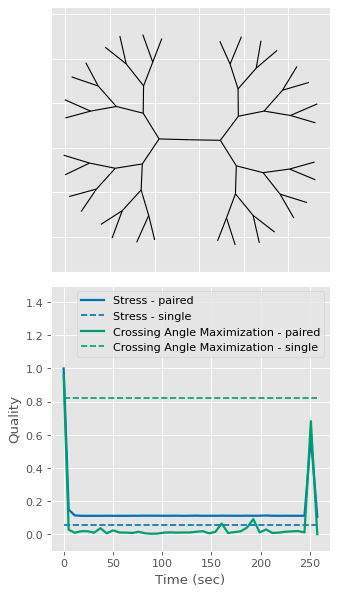}
    \includegraphics[width=.49\linewidth]{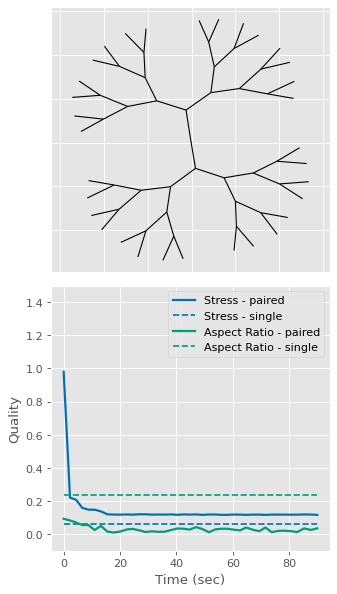}
    \caption{\blue{Better pairs}}
\end{subfigure}
\hfill
\begin{subfigure}[b]{0.32\linewidth}
    \centering
    \includegraphics[width=.48\linewidth]{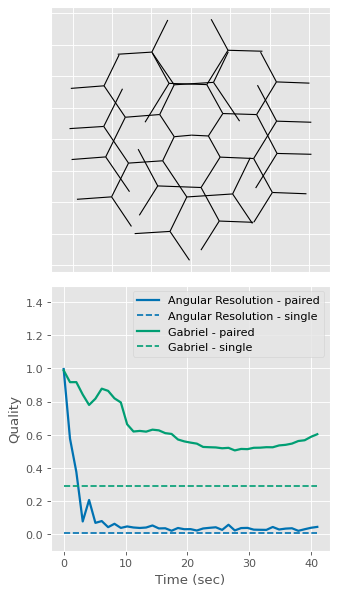}
    \includegraphics[width=.48\linewidth]{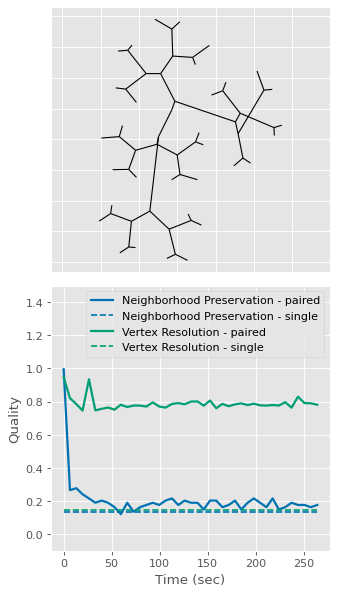}
    \caption{\blue{Worse pairs}}
\end{subfigure}
\caption{\blue{
We observed three types of interactions between criteria pairs: \textbf{(a)} compatible pairs can be optimized together; \textbf{(b)} better pairs do even better together than alone; \textbf{(c)} worse pairs are not fully compatible with each other. In the second row, stress is normalized to [0,1] based on its maximum value in each chart; angular resolution and vertex resolution are inverted $Q \mapsto 1-Q$ so that for all criteria smaller values are always better and the worst value is always 1.
}}
\label{fig:three-pairs}
\end{figure*}

\begin{figure*}[h]
\centering
\includegraphics[width=\textwidth]{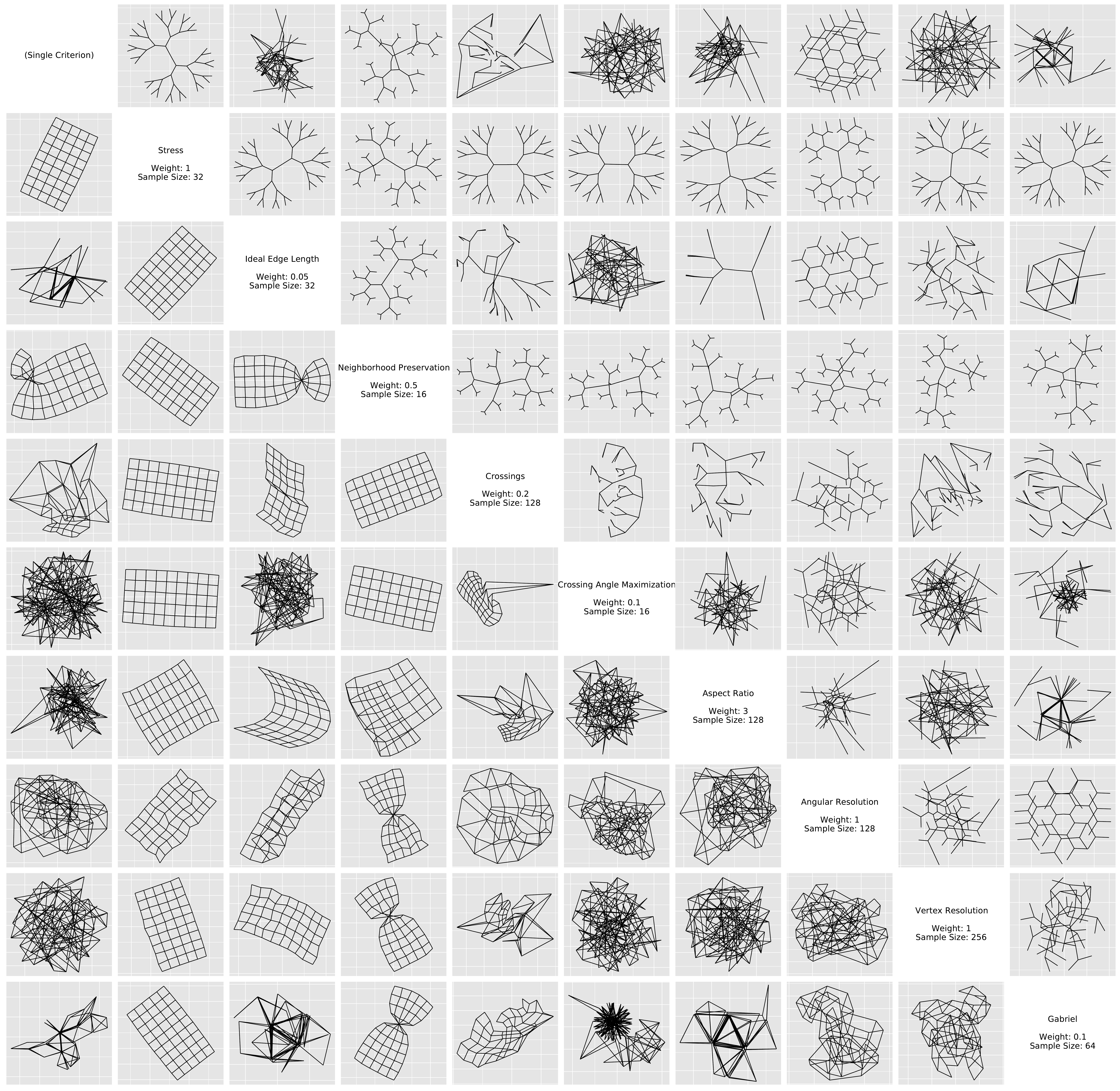}
\caption{
\blue{
Drawings of optimizing single or pair of criteria on a 6x10 grid with 60 nodes (lower left) or a 5-level balanced binary tree with 63 nodes (upper right). The sample size and weight for each criterion are shown in the diagonal entries of the figure.
}
}
\label{fig:analysis-pairs-drawings}
\end{figure*}

\subsection{Optimizing Layouts without Quality Decline}

Many optimization criteria tend to provide a drawing that has fewer crossings but cannot guarantee a crossing-free drawing. Stress is one such criterion, whose optimization does not guarantee crossing-free layout when graph is planar (or even a tree). One of the reasons for the popularity of stress-based layout methods is that they capture the underlying graph topology well. On the other hand, algorithms that are guaranteed to produce planar drawing (for planar graphs) are well known to dramatically distort the graph topology.

Our crossing minimization optimization is a soft constraint and does not guarantee a planar drawing when graph is planar. Hence, we provide an extra feature in our system that if we start with a planar drawing we can optimize any of the criteria above, while guaranteeing that no edge crossing will arise at any time. To do this, we add one additional test for every gradient descent step: for each proposed coordinate, we first check whether moving a node to this coordinate will introduce a crossing and if so, we do not update the coordinate. Using this method starting with an initial planar drawing, we can improve it, and provide a layout that is also planar and preserves the topology; see Fig.~\ref{fig:maintian_planarity}. 
This technique can be useful in other force-directed algorithms that do not guarantee crossing-free drawing when the initial layout is without crossings~\cite{fowler2012planar}.



We can generalize this idea for any graphs while optimizing any criterion. In the above scenario, we maintained the number of crossings of the layouts equal to zero since we started with a planar layout. If the graph is non-planar then we can update the coordinates in a similar way and the number of crossings in the progressive layouts will be decreasing. Similarly, we can consider other criteria, for example, the edge uniformity loss to decrease in the progressive layouts. If we are optimizing another criterion, for example, stress, there is no guarantee that edge uniformity will improve, see Fig.~\ref{fig:maintian_EU_triangular}. 
\blue{The general algorithm for safely update the layout with respect to any quality measure is described in Algorithm~\ref{alg:safe_update_2}.
Note that Algorithm~\ref{alg:safe_update_2} is applied in each SGD iteration of Algorithm \ref{alg:gd2}. 
Furthermore, it goes through all nodes in the graph and the quality measure is evaluated every time an intermediate layout is generated by a single node update. 
Hence, this approach does not  scale well to large graphs when the quality measure requires high computational overhead.
}

\section{Experimental Evaluation}
In this section, we assess the effectiveness and limitations of our approach. 
Since multiple criteria are not necessarily compatible with each other during optimization, we first identify all compatible pairs of criteria.
After identifying all compatible pairs, we hand-craft weighting schedules to optimize multiple criteria using \GDGD and compare our layouts from multicriteria optimization with classic drawing algorithms.
Finally, we analyze the runtime behavior and the impact of sample size in our approach.

\subsection{Interactions between Criteria}\label{sect:analysis-of-criteria-pairs}

\blue{We test the interactions between every pair of criteria using two regular graphs, a 6x10 grid (60 nodes) and a balanced binary tree with depth 5 (63 nodes).
Before dealing with pairs of criteria, we first test every single criterion to establish a lower bound of the quality measure.
In this section, we invert some quality measures (e.g., neighborhood preservation, and angular resolution) such that lower values are always better in all quality measures.
Then, we optimize every pair of criteria, monitor the quality measures of the pair over the course of training, and compare them with the corresponding lower bound found when optimizing each single criterion.}

\blue{As expected, we observe that all but one criterion, when optimized on their own, improve or maintain high quality during improvement iterations. The exception is crossing angle maximization, with a quality measure that depends on the worst crossing in the graph. The initial random layout usually has many crossings and maximizing crossing angles on its own (e.g., without also minimizing the number of crossings) does not necessarily lead to high-quality results.
Later we will see that optimizing other criteria together with crossing angle maximization helps. Further, when weight factors can be adjusted with a schedule, we recommend assigning positive weight to crossing angle maximization only at the later iterations.}

\blue{When optimizing pairs of aesthetic criteria, we see three types of pairs: compatible pairs, better pairs and worse pairs. Fig.~\ref{fig:three-pairs} shows an example for each of the three cases.
Most pairs are compatible pairs. 
For example, stress minimization is compatible with most other drawing aesthetics, as the qualities of both optimization goals can improve over time and they both achieve their lower bounds. 
Some pairs of criteria even do better together than alone. 
For example, crossing angle maximization together with stress minimization leads to better results than just crossing angle maximization, confirming the results of Huang et al.~\cite{huang2013}.
A few pairs of criteria are not fully compatible, leading to worse joint optimization. 
For example, when simultaneously optimizing vertex resolution and angular resolution, neither value can reach their corresponding lower bound.}

\begin{figure*}[thbp]
\centering
\includegraphics[width=\textwidth]{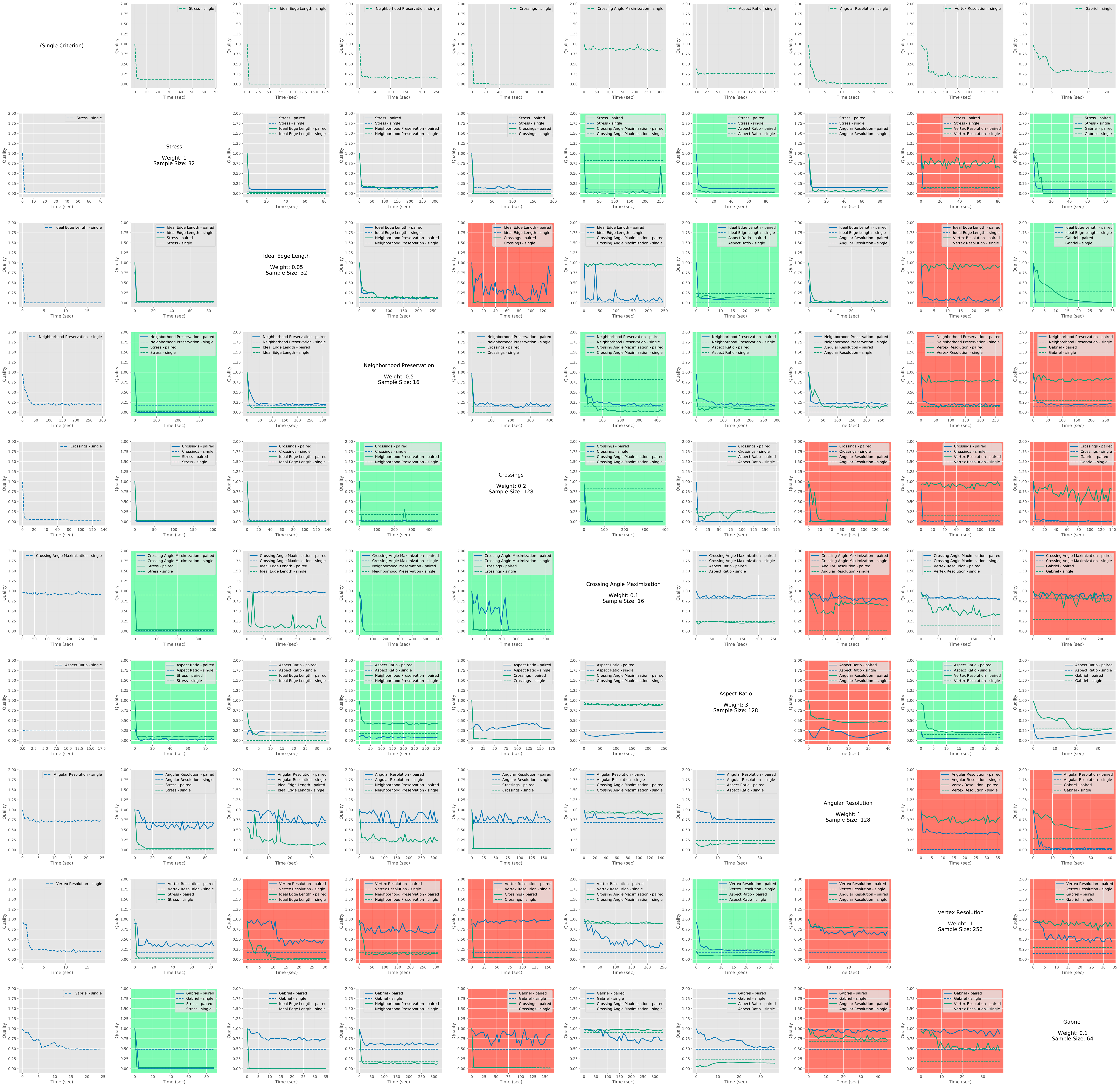}
\caption{
\blue{
Learning curves of optimizing pairs of criteria on a 6x10 grid with 60 nodes (lower left) or a 5-level balanced binary tree with 63 nodes (upper right). Better pairs are highlighted in green; worse pairs are highlighted in red. The sample size and weight for each criterion are shown in the diagonal entries of the figure. 
In this figure, some quality measures (neighborhood preservation, aspect ratio, angular resolution, vertex resolution and gabriel) are inverted $Q \mapsto 1-Q$, some (stress and ideal edge length) are normalized to the [0,1] interval so that among all criteria smaller values are always better and the worst value is always 1.
}
}
\label{fig:analysis-pairs-learning-curves}
\end{figure*}

\blue{Out of all 36 pairs, we find we find 20 compatible pairs, 9 better pairs and 7 worse pairs for the $6 \times 10$ grid; for the binary tree with depth 5, we find 
13 compatible pairs, 9 better pairs and 14 worse pairs. 
Comparing the compatibility between the two graphs, we note that all worse pairs in the grid are also worse pairs in the tree, and most of the better pairs and compatible pairs are shared between two graphs.
See 
the additional figures
(Fig.~\ref{fig:analysis-pairs-drawings} and \ref{fig:analysis-pairs-learning-curves}) 
for the drawings and quality curves of all criteria pairs and singletons.
It is worth noting that the compatibility of criteria also depends on the specific choice of weight factors. 
For example, having a dominating criterion by assigning a large weight to it can deteriorate the quality of the other.
In this analysis, we assign a fixed weight factor (and sample size) to each criterion so that every pair yields a reasonable outcome.}
%
%

\begin{figure*}[thbp]
\centering
\includegraphics[width=\textwidth,height=0.80\textheight,keepaspectratio]{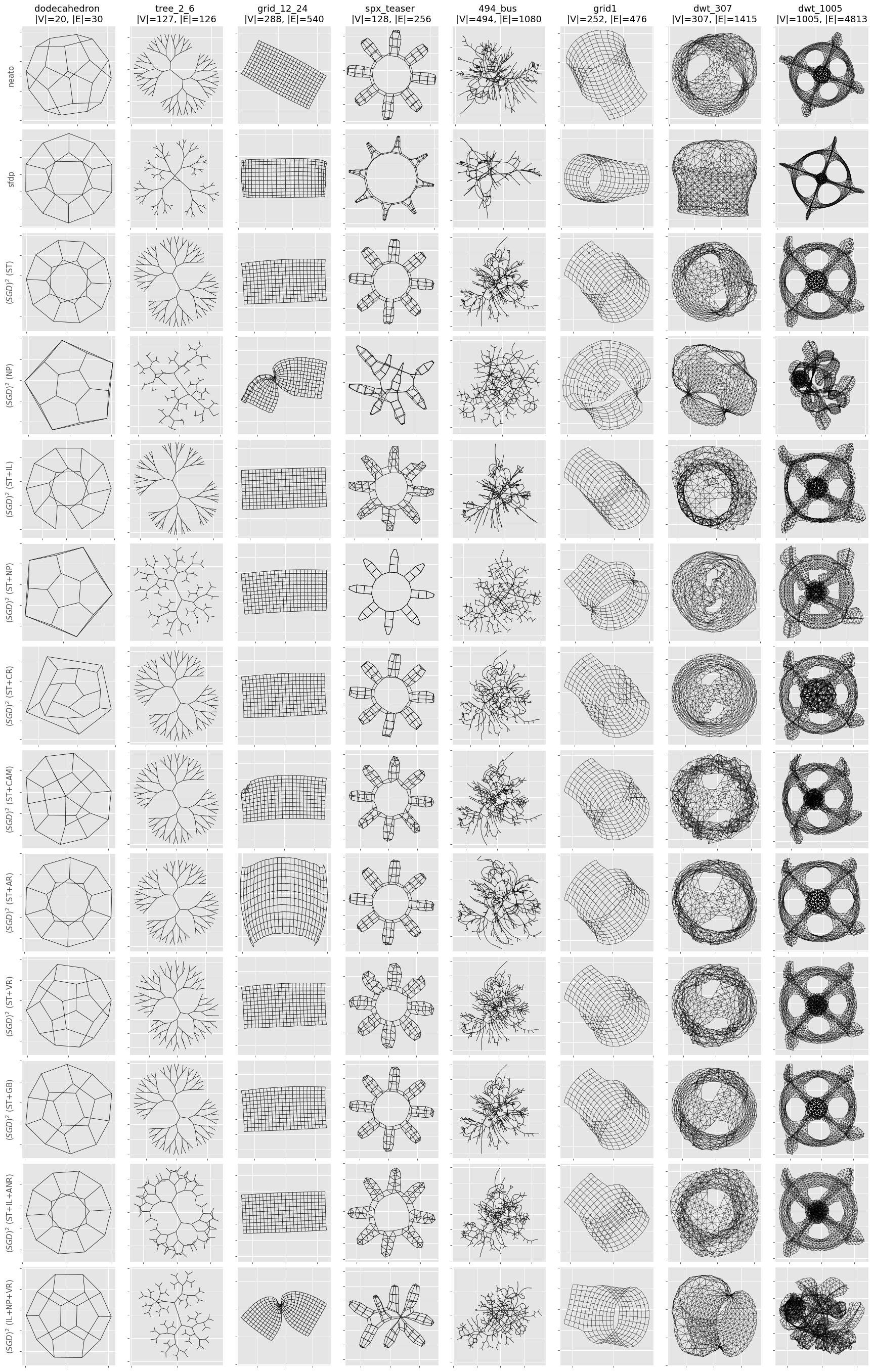}
\caption{
\blue{
Distinctive drawings from different algorithms: we compare the drawings of neato, sfdp and \GDGD that optimizes selected subsets of the 9 criteria: 
stress (\texttt{ST}), 
ideal edge length (\texttt{IL}), 
neighborhood preservation (\texttt{NP}), 
crossings (\texttt{CR}), 
crossing angle maximization (\texttt{CAM}), 
aspect ratio (\texttt{AR}),
angular resolution (\texttt{ANR}),
vertex resolution (\texttt{VR}), and
gabriel (\texttt{GB})
on 8 graphs.
}
}
\label{fig:all-drawings-partial}
\end{figure*}

\begin{table*}[thbp]
\begin{tabular}{l|rrrrrrrr}
\hline
 methods \textbackslash~graphs & dodecahedron   & tree-2-6       & grid-12-24     & spx-teaser     & 494-bus        & grid1          & dwt-307        & dwt-1005       \\
\hline
 neato                         & 0.081          & \textbf{0.078} & \textbf{0.013} & \textbf{0.027} & 0.076          & \textbf{0.062} & 0.083          & \textbf{0.022} \\
 sfdp                          & 0.080 & 0.133          & 0.024          & 0.052          & 0.099          & 0.071          & \textbf{0.081} & 0.029          \\
 \GDGD (ST)                      & \textbf{0.079} & \textbf{0.078} & \textbf{0.013} & \textbf{0.027} & \textbf{0.071} & \textbf{0.062} & 0.082          & \textbf{0.022} \\
 \GDGD (NP)                      & 0.188          & 0.233          & 0.065          & 0.241          & 0.825          & 0.149          & 0.279          & 0.275          \\
 \GDGD (ST+IL)                   & 0.107          & 0.100          & 0.033          & 0.054          & 0.090          & 0.100          & 0.119          & 0.043          \\
 \GDGD (ST+NP)                   & 0.188          & 0.106          & \textbf{0.013} & 0.051          & 0.739          & 0.080          & 0.178          & 0.059          \\
 \GDGD (ST+CR)                   & 0.190          & \textbf{0.078} & \textbf{0.013} & 0.028          & 0.079          & 0.073          & 0.091          & 0.045          \\
 \GDGD (ST+CAM)                  & 0.099          & \textbf{0.078} & 0.015          & 0.029          & 0.075          & 0.063          & 0.094          & 0.029          \\
 \GDGD (ST+AR)                   & 0.080 & 0.081          & 0.055          & \textbf{0.027} & 0.075          & 0.067          & 0.084          & 0.023          \\
 \GDGD (ST+VR)                   & 0.083          & 0.080          & \textbf{0.013} & 0.032          & 0.073          & 0.063          & 0.088          & 0.023          \\
 \GDGD (ST+GB)                   & 0.080 & \textbf{0.078} & \textbf{0.013} & \textbf{0.027} & \textbf{0.071} & \textbf{0.062} & 0.083          & \textbf{0.022} \\
 \GDGD (ST+IL+ANR)               & 0.089          & 0.098          & 0.020          & 0.036          & 0.080          & 0.084          & 0.096          & 0.027          \\
 \GDGD (IL+NP+VR)                & 0.083          & 0.119          & 0.061          & 0.134          & 0.208          & 0.068          & 0.110          & 0.205          \\
\hline
\end{tabular}
\caption{\blue{Quality Measures of Stress (ST)}}
\label{tab:quality-table-stress}
\end{table*}

\begin{table*}[thbp]
\begin{tabular}{l|rrrrrrrr}
\hline
 methods \textbackslash~graphs & dodecahedron   & tree-2-6       & grid-12-24     & spx-teaser     & 494-bus        & grid1          & dwt-307        & dwt-1005       \\
\hline
 neato                         & 0.028          & 0.005          & \textbf{0.002} & 0.099          & 0.484          & 0.026          & 0.299          & 0.280          \\
 sfdp                          & 0.018          & 0.143          & 0.051          & 0.271          & 0.585          & 0.052          & 0.279          & 0.346          \\
 \GDGD (ST)                      & 0.062          & 0.103          & 0.053          & 0.114          & 0.526          & 0.111          & 0.320          & 0.293          \\
 \GDGD (NP)                      & 0.679          & 0.867          & 0.351          & 0.884          & 3.348          & 0.649          & 1.083          & 1.312          \\
 \GDGD (ST+IL)                   & \textbf{0.004} & \textbf{0.003} & \textbf{0.002} & \textbf{0.009} & \textbf{0.462} & \textbf{0.002} & \textbf{0.249} & \textbf{0.245} \\
 \GDGD (ST+NP)                   & 0.679          & 0.227          & 0.056          & 0.434          & 1.608          & 0.418          & 0.749          & 0.519          \\
 \GDGD (ST+CR)                   & 0.517          & 0.103          & 0.053          & 0.139          & 0.593          & 0.205          & 0.394          & 0.482          \\
 \GDGD (ST+CAM)                  & 0.109          & 0.103          & 0.058          & 0.132          & 0.595          & 0.141          & 0.497          & 0.368          \\
 \GDGD (ST+AR)                   & 0.061          & 0.112          & 0.203          & 0.114          & 0.533          & 0.148          & 0.334          & 0.302          \\
 \GDGD (ST+VR)                   & 0.083          & 0.257          & 0.067          & 0.159          & 0.569          & 0.155          & 0.399          & 0.309          \\
 \GDGD (ST+GB)                   & 0.061          & 0.098          & 0.053          & 0.092          & 0.521          & 0.111          & 0.338          & 0.291          \\
 \GDGD (ST+IL+ANR)               & 0.016          & 0.027          & 0.012          & 0.029          & 0.486          & 0.027          & 0.277          & 0.268          \\
 \GDGD (IL+NP+VR)                & 0.082          & 0.178          & 0.116          & 0.242          & 0.702          & 0.147          & 0.388          & 0.549          \\
\hline
\end{tabular}
\caption{\blue{Quality Measures of Ideal Edge Length (IL)}}
\label{tab:quality-table-ideal_edge_length}
\end{table*}

\begin{table*}[thbp]
\begin{tabular}{l|rrrrrrrr}
\hline
 methods \textbackslash~graphs & dodecahedron   & tree-2-6       & grid-12-24     & spx-teaser     & 494-bus        & grid1          & dwt-307        & dwt-1005       \\
\hline
 neato                         & 0.723          & 0.718          & \textbf{0.000} & 0.474          & 0.846          & 0.558          & 0.699          & 0.545          \\
 sfdp                          & 0.571          & 0.592          & 0.063          & 0.533          & 0.750          & 0.651          & 0.584          & 0.653          \\
 \GDGD (ST)                      & 0.500          & 0.721          & \textbf{0.000} & 0.503          & 0.832          & 0.553          & 0.657          & 0.548          \\
 \GDGD (NP)                      & \textbf{0.400} & 0.225          & 0.276          & 0.487          & \textbf{0.659} & 0.480          & \textbf{0.428} & 0.516          \\
 \GDGD (ST+IL)                   & 0.667          & 0.701          & \textbf{0.000} & 0.518          & 0.867          & 0.674          & 0.766          & 0.665          \\
 \GDGD (ST+NP)                   & \textbf{0.400} & \textbf{0.194} & \textbf{0.000} & 0.467          & 0.664          & \textbf{0.347} & 0.472          & \textbf{0.441} \\
 \GDGD (ST+CR)                   & 0.481          & 0.714          & \textbf{0.000} & \textbf{0.420} & 0.765          & 0.532          & 0.582          & 0.567          \\
 \GDGD (ST+CAM)                  & 0.681          & 0.721          & 0.042          & 0.531          & 0.851          & 0.552          & 0.763          & 0.631          \\
 \GDGD (ST+AR)                   & 0.500          & 0.727          & 0.418          & 0.492          & 0.842          & 0.560          & 0.757          & 0.584          \\
 \GDGD (ST+VR)                   & 0.750          & 0.708          & \textbf{0.000} & 0.609          & 0.834          & 0.527          & 0.709          & 0.570          \\
 \GDGD (ST+GB)                   & 0.696          & 0.727          & \textbf{0.000} & 0.566          & 0.831          & 0.555          & 0.702          & 0.547          \\
 \GDGD (ST+IL+ANR)               & 0.500          & 0.749          & \textbf{0.000} & 0.539          & 0.823          & 0.622          & 0.705          & 0.520          \\
 \GDGD (IL+NP+VR)                & 0.636          & \textbf{0.194} & 0.165          & 0.472          & 0.713          & 0.349          & 0.497          & 0.717          \\
\hline
\end{tabular}
\caption{\blue{Quality Measures of Neighborhood Preservation (NP)}}
\label{tab:quality-table-neighborhood_preservation}
\end{table*}

\begin{table*}[thbp]
\begin{tabular}{l|rrrrrrrr}
\hline
 methods \textbackslash~graphs & dodecahedron & tree-2-6   & grid-12-24 & spx-teaser  & 494-bus      & grid1       & dwt-307      & dwt-1005      \\
\hline
 neato                         & 8            & 1          & \textbf{0} & 80          & 274          & 134         & 1896         & 8001          \\
 sfdp                          & 9            & 2          & \textbf{0} & \textbf{72} & 141          & 109         & 1651         & \textbf{5447} \\
 \GDGD (ST)                      & 10           & \textbf{0} & \textbf{0} & 73          & 283          & 133         & 1749         & 8367          \\
 \GDGD (NP)                      & 10           & 5          & 109        & 155         & 216          & \textbf{80} & 1251         & 10736         \\
 \GDGD (ST+IL)                   & 10           & \textbf{0} & \textbf{0} & 73          & 325          & 151         & 3924         & 13094         \\
 \GDGD (ST+NP)                   & 6            & \textbf{0} & \textbf{0} & 111         & \textbf{131} & 94          & \textbf{265} & 6932          \\
 \GDGD (ST+CR)                   & \textbf{0}   & \textbf{0} & \textbf{0} & 73          & 187          & 133         & 347          & 8827          \\
 \GDGD (ST+CAM)                  & 7            & \textbf{0} & 9          & 84          & 430          & 158         & 4673         & 12972         \\
 \GDGD (ST+AR)                   & 10           & \textbf{0} & 2          & \textbf{72} & 303          & 142         & 2836         & 8628          \\
 \GDGD (ST+VR)                   & 7            & \textbf{0} & \textbf{0} & 79          & 282          & 134         & 2983         & 8874          \\
 \GDGD (ST+GB)                   & 6            & \textbf{0} & \textbf{0} & 75          & 264          & 134         & 2034         & 8305          \\
 \GDGD (ST+IL+ANR)               & 10           & 24         & \textbf{0} & \textbf{72} & 302          & 171         & 2793         & 8241          \\
 \GDGD (IL+NP+VR)                & 6            & 2          & 96         & 105         & 173          & 103         & 1333         & 18276         \\
\hline
\end{tabular}
\caption{\blue{Quality Measures of Crossings (CR)}}
\label{tab:quality-table-crossings}
\end{table*}

\begin{table*}[thbp]
\begin{tabular}{l|rrrrrrrr}
\hline
 methods \textbackslash~graphs & dodecahedron   & tree-2-6       & grid-12-24     & spx-teaser     & 494-bus        & grid1          & dwt-307        & dwt-1005       \\
\hline
 neato                         & 0.254          & 0.419          & \textbf{0.000} & 0.786          & 0.956          & 0.927          & 0.970          & 0.999          \\
 sfdp                          & 0.622          & 0.235          & \textbf{0.000} & 0.806          & 0.972          & 0.970          & 0.963          & 1.000          \\
 \GDGD (ST)                      & 0.601          & \textbf{0.000} & \textbf{0.000} & 0.975          & 0.923          & 0.946          & 0.983          & 0.990          \\
 \GDGD (NP)                      & 0.969          & 0.808          & 0.945          & 0.999          & 0.913          & 0.939          & 0.988          & 0.992          \\
 \GDGD (ST+IL)                   & 0.602          & \textbf{0.000} & \textbf{0.000} & 0.575          & 0.970          & 0.837          & 0.989          & 0.998          \\
 \GDGD (ST+NP)                   & 0.969          & \textbf{0.000} & \textbf{0.000} & 1.000          & \textbf{0.838} & 0.864          & 0.958          & 0.992          \\
 \GDGD (ST+CR)                   & \textbf{0.000} & \textbf{0.000} & \textbf{0.000} & 0.940          & 0.885          & 0.889          & 0.955          & \textbf{0.984} \\
 \GDGD (ST+CAM)                  & \textbf{0.000} & \textbf{0.000} & 0.810          & 0.646          & 0.937          & \textbf{0.622} & 0.973          & 0.993          \\
 \GDGD (ST+AR)                   & 0.605          & \textbf{0.000} & 0.757          & 0.712          & 0.985          & 0.965          & 0.978          & 0.995          \\
 \GDGD (ST+VR)                   & 0.164          & \textbf{0.000} & \textbf{0.000} & 0.550          & 0.943          & 0.883          & 0.975          & 0.995          \\
 \GDGD (ST+GB)                   & 0.524          & \textbf{0.000} & \textbf{0.000} & 0.834          & 0.945          & 0.937          & 0.992          & 0.991          \\
 \GDGD (ST+IL+ANR)               & 0.603          & 0.725          & \textbf{0.000} & \textbf{0.421} & 0.935          & 0.961          & 0.973          & 0.997          \\
 \GDGD (IL+NP+VR)                & 0.323          & 0.143          & 0.954          & 0.963          & 0.913          & 0.631          & \textbf{0.926} & 0.991          \\
\hline
\end{tabular}
\caption{\blue{Quality Measures of Crossing Angle Maximization (CAM)}}
\label{tab:quality-table-crossing_angle_maximization}
\end{table*}

\begin{table*}[thbp]
\begin{tabular}{l|rrrrrrrr}
\hline
 methods \textbackslash~graphs & dodecahedron   & tree-2-6       & grid-12-24     & spx-teaser     & 494-bus        & grid1          & dwt-307        & dwt-1005       \\
\hline
 neato                         & 0.062          & 0.145          & 0.483          & 0.010          & 0.143          & 0.314          & 0.065          & \textbf{0.012} \\
 sfdp                          & 0.068          & 0.084          & 0.536          & 0.010          & 0.192          & 0.452          & 0.095          & 0.018          \\
 \GDGD (ST)                      & 0.049          & 0.150          & 0.470          & \textbf{0.002} & 0.155          & 0.315          & 0.071          & 0.094          \\
 \GDGD (NP)                      & \textbf{0.047} & 0.124          & 0.481          & 0.176          & 0.162          & \textbf{0.049} & 0.109          & 0.282          \\
 \GDGD (ST+IL)                   & \textbf{0.048} & 0.169          & 0.507          & 0.018          & 0.160          & 0.470          & 0.157          & 0.045          \\
 \GDGD (ST+NP)                   & 0.049          & 0.126          & 0.479          & \textbf{0.002} & 0.192          & 0.209          & 0.077          & 0.101          \\
 \GDGD (ST+CR)                   & 0.134          & 0.149          & 0.473          & 0.028          & 0.136          & 0.368          & 0.030          & 0.079          \\
 \GDGD (ST+CAM)                  & 0.068          & 0.149          & 0.449          & 0.009          & 0.188          & 0.288          & \textbf{0.028} & 0.141          \\
 \GDGD (ST+AR)                   & \textbf{0.047} & \textbf{0.017} & \textbf{0.048} & 0.008          & \textbf{0.118} & 0.154          & 0.043          & 0.057          \\
 \GDGD (ST+VR)                   & 0.068          & 0.135          & 0.466          & 0.009          & 0.152          & 0.300          & 0.044          & 0.091          \\
 \GDGD (ST+GB)                   & 0.069          & 0.154          & 0.470          & 0.013          & 0.155          & 0.318          & 0.037          & 0.093          \\
 \GDGD (ST+IL+ANR)               & \textbf{0.048} & 0.197          & 0.508          & 0.045          & 0.178          & 0.269          & 0.058          & 0.084          \\
 \GDGD (IL+NP+VR)                & 0.101          & 0.231          & 0.455          & 0.242          & 0.359          & 0.282          & 0.074          & 0.106          \\
\hline
\end{tabular}
\caption{\blue{Quality Measures of Aspect Ratio (AR)}}
\label{tab:quality-table-aspect_ratio}
\end{table*}

\begin{table*}[thbp]
\begin{tabular}{l|rrrrrrrr}
\hline
 methods \textbackslash~graphs & dodecahedron   & tree-2-6       & grid-12-24     & spx-teaser     & 494-bus        & grid1          & dwt-307        & dwt-1005       \\
\hline
 neato                         & 0.753          & 0.749          & 0.525          & 0.999          & 0.998          & 1.000          & 1.000          & \textbf{1.000} \\
 sfdp                          & 0.459          & 0.908          & 0.547          & 0.996          & 0.999          & 0.996          & 1.000          & \textbf{1.000} \\
 \GDGD (ST)                      & \textbf{0.401} & 0.750          & 0.524          & 0.989          & \textbf{0.993} & 0.941          & 0.999          & \textbf{1.000} \\
 \GDGD (NP)                      & 0.928          & 0.976          & 0.994          & 0.996          & 1.000          & 0.984          & 1.000          & \textbf{1.000} \\
 \GDGD (ST+IL)                   & 0.517          & 0.844          & 0.497          & 0.939          & 0.999          & 0.932          & 0.999          & \textbf{1.000} \\
 \GDGD (ST+NP)                   & 0.952          & 0.466          & \textbf{0.487} & 1.000          & 0.999          & 0.998          & 0.999          & \textbf{1.000} \\
 \GDGD (ST+CR)                   & 0.746          & 0.769          & 0.526          & 0.996          & 0.998          & 0.969          & \textbf{0.998} & \textbf{0.999} \\
 \GDGD (ST+CAM)                  & 1.000          & 0.745          & 0.871          & 0.913          & 1.000          & 0.999          & 1.000          & \textbf{0.999} \\
 \GDGD (ST+AR)                   & 0.403          & 0.779          & 0.991          & 0.963          & 0.998          & 0.994          & 0.999          & \textbf{1.000} \\
 \GDGD (ST+VR)                   & 0.760          & 0.979          & 0.540          & 0.936          & 1.000          & 0.914          & 1.000          & \textbf{1.000} \\
 \GDGD (ST+GB)                   & 0.573          & 0.750          & 0.528          & \textbf{0.868} & 1.000          & 0.903          & 1.000          & \textbf{1.000} \\
 \GDGD (ST+IL+ANR)               & \textbf{0.401} & \textbf{0.327} & 0.528          & 0.990          & 1.000          & \textbf{0.740} & \textbf{0.999} & \textbf{0.999} \\
 \GDGD (IL+NP+VR)                & 0.677          & 0.418          & 0.974          & 0.995          & 0.999          & 0.823          & \textbf{0.999} & \textbf{1.000} \\
\hline
\end{tabular}
\caption{\blue{Quality Measures of Angular Resolution (ANR)}}
\label{tab:quality-table-angular_resolution}
\end{table*}

\begin{table*}[thbp]
\begin{tabular}{l|rrrrrrrr}
\hline
 methods \textbackslash~graphs & dodecahedron   & tree-2-6       & grid-12-24     & spx-teaser     & 494-bus        & grid1          & dwt-307        & dwt-1005       \\
\hline
 neato                         & 0.637          & 0.735          & 0.362          & 0.765          & 0.948          & 0.780          & 0.870          & \textbf{0.931} \\
 sfdp                          & 0.391          & 0.594          & 0.807          & 0.825          & 0.951          & 0.877          & 0.859          & 0.967          \\
 \GDGD (ST)                      & 0.269          & 0.727          & 0.368          & 0.663          & 0.991          & 0.798          & 0.869          & 0.976          \\
 \GDGD (NP)                      & 1.000          & 0.875          & 0.902          & 0.992          & 0.986          & 0.814          & 0.827          & 0.987          \\
 \GDGD (ST+IL)                   & 0.334          & 0.705          & \textbf{0.354} & 0.788          & 0.993          & 0.973          & 0.994          & 0.993          \\
 \GDGD (ST+NP)                   & 1.000          & 0.688          & 0.454          & 1.000          & 0.989          & 0.791          & 0.862          & 0.977          \\
 \GDGD (ST+CR)                   & 0.570          & 0.762          & 0.379          & 0.789          & 0.954          & 0.908          & 0.785          & 0.950          \\
 \GDGD (ST+CAM)                  & 0.996          & 0.715          & 0.839          & 0.789          & 0.974          & 0.856          & 0.967          & 0.959          \\
 \GDGD (ST+AR)                   & 0.278          & 0.682          & 0.633          & 0.698          & 0.938          & 0.802          & 0.963          & 0.991          \\
 \GDGD (ST+VR)                   & \textbf{0.165} & \textbf{0.266} & 0.388          & \textbf{0.407} & \textbf{0.754} & \textbf{0.590} & \textbf{0.617} & 0.945          \\
 \GDGD (ST+GB)                   & 0.529          & 0.737          & 0.365          & 0.630          & 0.895          & 0.858          & 0.913          & 0.986          \\
 \GDGD (ST+IL+ANR)               & 0.332          & 0.959          & 0.392          & 0.873          & 0.961          & 0.959          & 0.931          & 0.984          \\
 \GDGD (IL+NP+VR)                & 0.355          & 0.615          & 0.795          & 0.744          & 0.831          & \textbf{0.590} & 0.751          & 0.993          \\
\hline
\end{tabular}
\caption{\blue{Quality Measures of Vertex Resolution (VR)}}
\label{tab:quality-table-vertex_resolution}
\end{table*}

\begin{table*}[thbp]
\begin{tabular}{l|rrrrrrrr}
\hline
 methods \textbackslash~graphs & dodecahedron   & tree-2-6       & grid-12-24     & spx-teaser     & 494-bus        & grid1          & dwt-307        & dwt-1005       \\
\hline
 neato                         & 0.429          & 0.595          & \textbf{0.000} & 0.933          & \textbf{1.000} & 0.920          & \textbf{1.000} & \textbf{1.000} \\
 sfdp                          & 0.860          & 0.806          & 0.148          & 0.758          & \textbf{1.000} & 0.967          & \textbf{1.000} & \textbf{1.000} \\
 \GDGD (ST)                      & 0.677          & 0.130          & \textbf{0.000} & 0.795          & \textbf{1.000} & 0.963          & \textbf{1.000} & \textbf{1.000} \\
 \GDGD (NP)                      & 0.951          & 0.954          & 0.973          & 0.920          & \textbf{1.000} & 0.903          & \textbf{1.000} & \textbf{1.000} \\
 \GDGD (ST+IL)                   & 0.679          & \textbf{0.000} & \textbf{0.000} & 0.834          & \textbf{1.000} & 0.860          & \textbf{1.000} & \textbf{1.000} \\
 \GDGD (ST+NP)                   & 0.951          & 0.658          & \textbf{0.000} & \textbf{0.039} & \textbf{1.000} & 0.916          & \textbf{1.000} & \textbf{1.000} \\
 \GDGD (ST+CR)                   & 0.704          & 0.205          & \textbf{0.000} & 0.967          & \textbf{1.000} & 0.988          & \textbf{1.000} & \textbf{1.000} \\
 \GDGD (ST+CAM)                  & 0.695          & 0.199          & 0.799          & 0.836          & \textbf{1.000} & 0.905          & \textbf{1.000} & \textbf{1.000} \\
 \GDGD (ST+AR)                   & 0.678          & 0.396          & 0.887          & 0.903          & \textbf{1.000} & 0.958          & \textbf{1.000} & \textbf{1.000} \\
 \GDGD (ST+VR)                   & 0.437          & 0.957          & \textbf{0.000} & 0.862          & \textbf{1.000} & 0.987          & \textbf{1.000} & \textbf{1.000} \\
 \GDGD (ST+GB)                   & \textbf{0.368} & 0.036          & \textbf{0.000} & 0.664          & \textbf{1.000} & \textbf{0.791} & \textbf{1.000} & \textbf{1.000} \\
 \GDGD (ST+IL+ANR)               & 0.681          & 0.745          & \textbf{0.000} & 0.630          & \textbf{1.000} & 0.967          & \textbf{1.000} & \textbf{1.000} \\
 \GDGD (IL+NP+VR)                & 0.421          & 0.800          & 0.982          & 0.926          & \textbf{1.000} & 0.972          & \textbf{1.000} & \textbf{1.000} \\
\hline
\end{tabular}
\caption{\blue{Quality Measures of Gabriel Property (GB)}}
\label{tab:quality-table-gabriel}
\end{table*}

\subsection{Quality Analysis}\label{sect:analysis-of-qualities}


\blue{We compare layouts obtained with \GDGD when optimizing different aesthetic goals to layouts obtained by  neato~\cite{ellson2001graphviz} and sfdp~\cite{ellson2001graphviz}, which are classic implementations of  stress-majorization and scalable force-directed methods. 
 Fig.~\ref{fig:all-drawings-partial} shows the layouts along with information about each graph. 
The graphs are chosen to represent a variety of classes such as trees, grids, regular shapes, and to also include real-world examples. 
In particular, the last four graphs in Fig.~\ref{fig:all-drawings-partial} are from the Sparse Matrix Collection~\cite{davis2011university} and are also used to evaluate stress minimization via SGD in~\cite{zheng2018graph}; see the supplementary materials for more layouts.
}

\blue{
Next, we evaluate each layout on 9 readability criteria: stress (\texttt{ST}), node resolution (\texttt{VR}), ideal edge lengths (\texttt{IL}), neighbor preservation (\texttt{NP}), crossings (\texttt{CR}), crossing angle (\texttt{CA}), angular resolution (\texttt{ANR}), aspect ratio (\texttt{AR}), and Gabriel graph property (\texttt{GB}). 
Our experiment utilizes 8 graphs and layouts computed by neato, sfdp, and 7 runs of \GDGD using various combinations of objectives.
\blue{
Tables~\ref{tab:quality-table-stress} to~\ref{tab:quality-table-gabriel} summarize the first 4 of the 9 quality measures for the layouts in Fig.~\ref{fig:all-drawings-partial}.
More combinations of criteria used for \GDGD and the remaining quality measures are included in the supplementary materials. 
The quality measure for crossings is the actual number of edge crossings in the layout. For all other criteria, we use the formulas defined in Section~\ref{sect:properties-and-measures}. All  quality measures produce values greater than or equal to zero: the lower the value the better the measure. In each column, the best score is bold.}
When optimizing via multicriteria \GDGD, we choose compatible pairs, better pairs, or compatible triples among the 9 criteria.
When optimizing incompatible pairs or triples, we fix the number of iterations in \GDGD, select and prioritize one criterion (or compatible pair) in an early stage of the training and postpone the others to the later stage.
For example, when simultaneously optimizing ideal edge length (IL), neighborhood preservation (NP) and vertex resolution (VR), we assign zero weight to VR and positive weights to IL and NP in the first half of the iterations. Then we gradually decrease the weights of IL and NP to 0 (by a smooth function that interpolates the highest and lowest weights) and increase the weight of VR in the second half of the iterations with a similar smooth growth function.
At each stage, we interpolate the two weight levels of each criterion $w_{start}$ and $w_{stop}$ between the start and stopping iterations $t_{start}$ to $t_{stop}$ by a scaled and translated smooth-step function $g(t)$: 
$$
g(t) = (w_{stop}-w_{start}) \cdot f(\frac{t-t_{start}}{t_{stop}-t_{start}}) + w_{start}
$$
where $f(x) = 3x^2-2x^3$ for $x \in [0,1]$ is typically called the (standard) smooth-step function.
}

\blue{
The experimental results confirm  that \GDGD yields better or comparable results for most quality measures and on most graphs.
We do note that some criteria (e.g., CR and GB) are harder to optimize on real-world large graphs; improving the performance on such tough  criteria is natural direction for future work.}

\begin{figure*}[t]
\centering
\includegraphics[width=\textwidth]{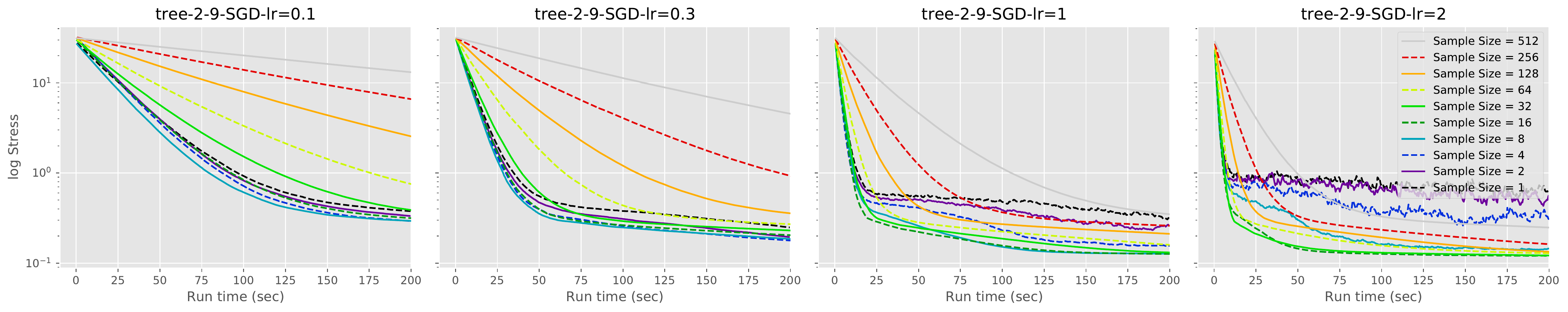}
\caption{\blue{Optimal sample size for stress minimization depends on the learning rate.
(\textbf{two figures on the left}) a smaller sample size (e.g., 4 or 8) gives faster convergence when the learning rate is relatively small; (\textbf{two figures on the right}) a medium sample size (e.g., 16 or 32) benefits from larger learning rate when training with smaller sample sizes becomes less stable and may give even faster convergence rate than using smaller sample size in the low-learning-rate cases.}}
\label{fig:analysis-of-sample-size}
\end{figure*}

\subsection{Analysis of Sample Size}\label{sect:analysis-of-sample-size}
\blue{
In this section we analyze the impact of sample size on the convergence rate of \GDGD.
In deep learning, models trained with different sample sizes can converge to different types of minima; e.g., smaller sample tend to lead to a better generalization\cite{keskar2016large}.
In \GDGD, smaller sample size usually results in faster run time \textit{per iteration} but does not necessarily yield faster \textit{per-second} convergence. 
As described in Section~\ref{sect:properties-and-measures}, we use different sampling strategies and sample sizes for each readability criterion.
Consider, for example, stress minimization and how optimal sample size closely depends on other factors in the optimization. 
In other words, there is no ``one size fits all'' sample size. 
In particular, for any given graph, the optimal sample size depends on the learning rate of the SGD algorithm.
Fig.~\ref{fig:analysis-of-sample-size} shows the quality (i.e. stress) of layouts for a binary tree with 9 levels (1023 nodes) as a function of total run time of the \GDGD algorithm.
In each plot, we visualize the convergence of the algorithm under a fixed learning rate with various sample sizes.
When the learning rate is small (the two plots on the left of Fig.~\ref{fig:analysis-of-sample-size}), we observe that smaller sample sizes (e.g., 4 or 8) converge faster.
In contrast, a medium sample size (e.g., 16 or 32) can benefit from larger learning rates (the two plots on the right of Fig.~\ref{fig:analysis-of-sample-size}) and converge faster than any cases that use a smaller learning rate.
Moreover, when using a large learning rate, the training with a smaller sample size becomes less stable (due to the high variance of gradients, see the rightmost plot in Fig.~\ref{fig:analysis-of-sample-size}).
We illustrate this observation on the binary tree graph with 9 levels (1023 nodes) and have the same observation on other trees and grids with various sizes.
Moreover, we observed that this interplay between sample size and learning rate on convergence rate is less obvious in variants of the SGD algorithm.
When replacing the SGD with some of its variants (e.g., AdaDelta~\cite{zeiler2012adadelta}, RMSProp~\cite{Tieleman2012} or ADAM~\cite{kingma2014adam}) that takes adaptive step size based on the gradient of previous steps, the convergence rate of stress minimization becomes less sensitive to sample size or learning rate.
}

\begin{figure}[t]
\centering
\includegraphics[trim={20 0 0 0}, clip, width=0.49\columnwidth]{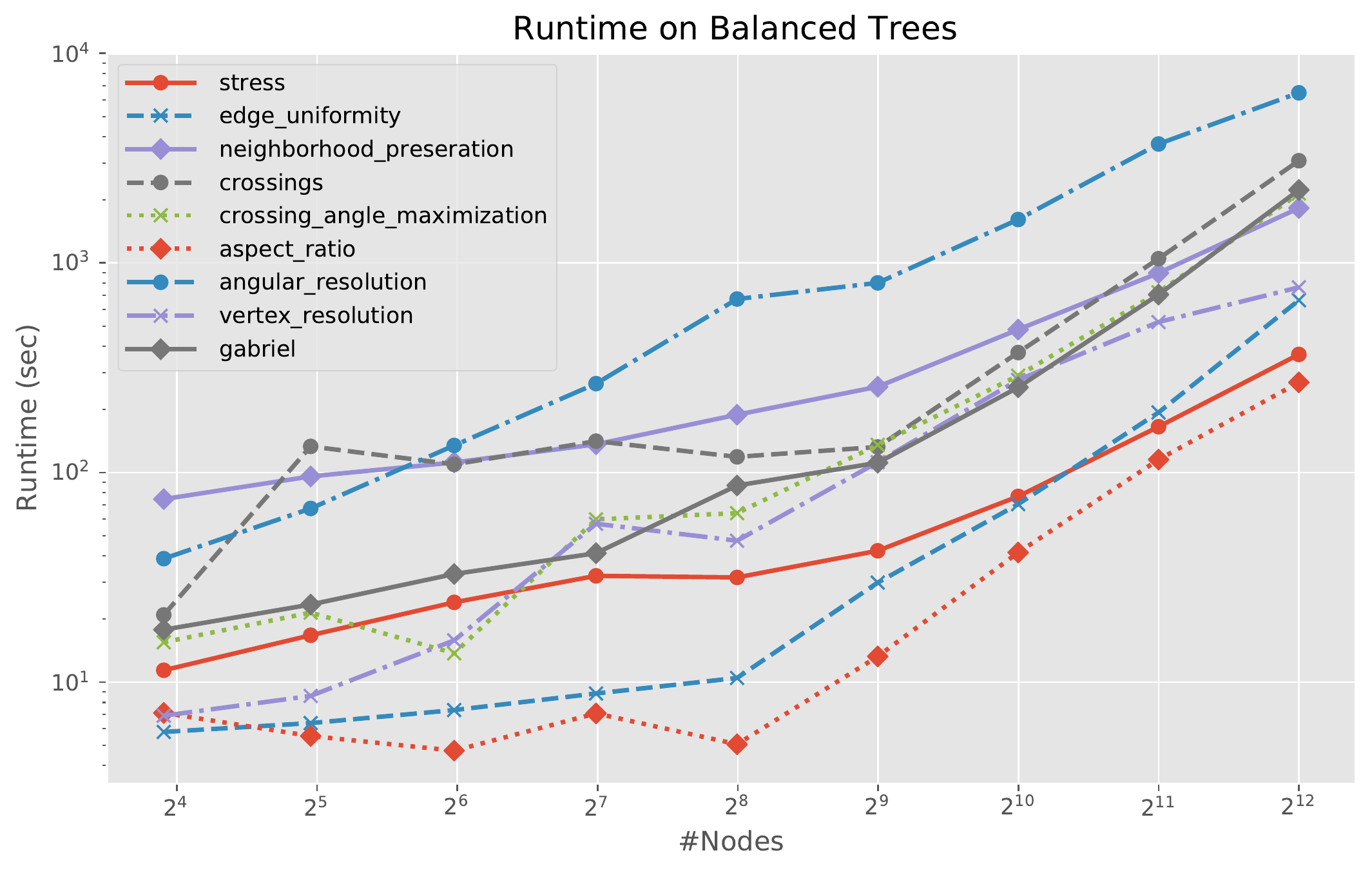}
\includegraphics[trim={20 0 0 0}, clip, width=0.49\columnwidth]{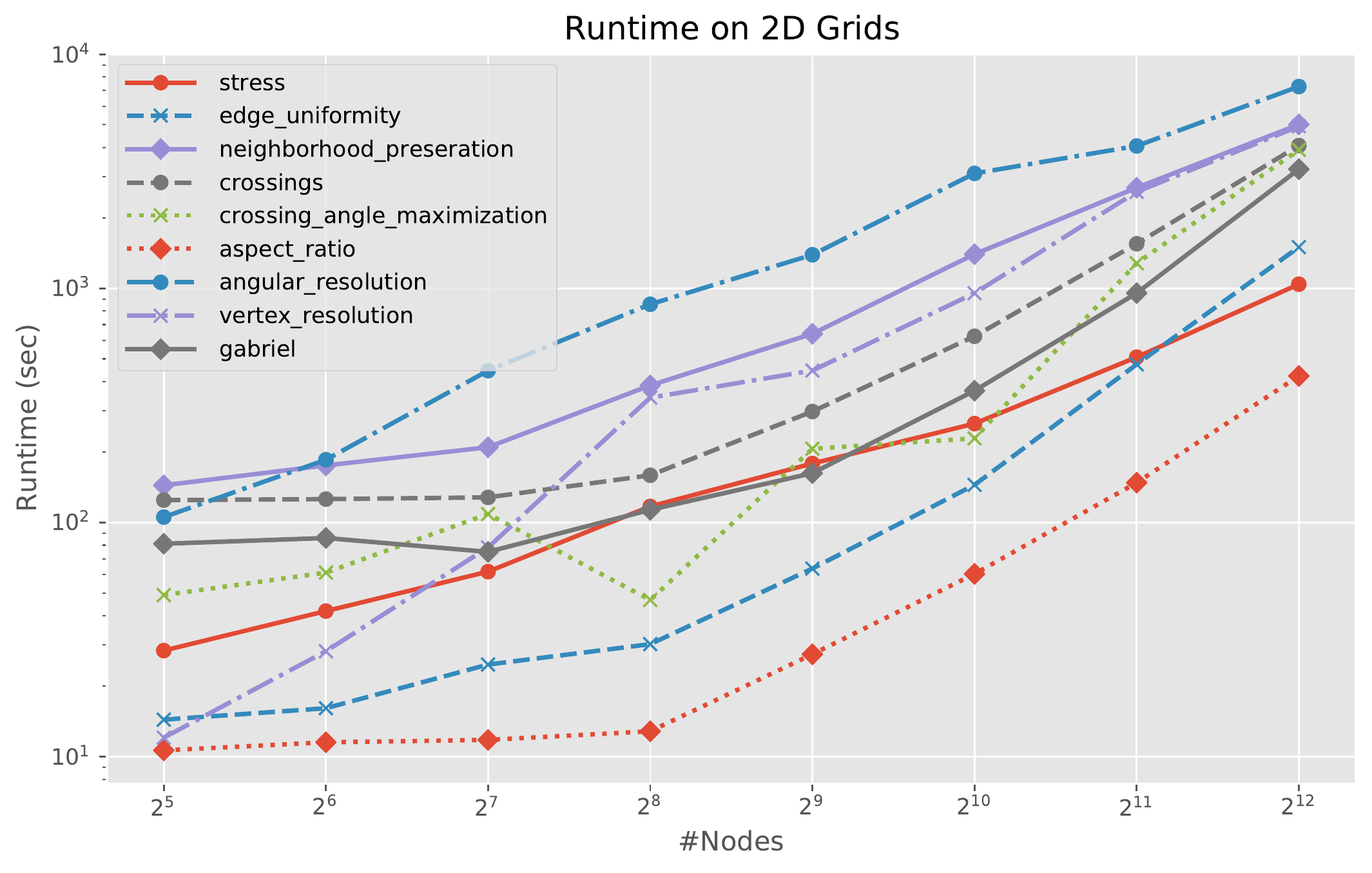}
\caption{
Runtime of balanced trees (top) and 2D grids (bottom). 
The plots have log scales on both x and y axes.
}
\label{fig:runtime}
\end{figure}

\subsection{Analysis of Run Time}
To test the scalability of our method, we test the runtime of our method on larger graphs. 
We tested our code on a MacBook Pro with a 2.9 GHz Dual-Core Intel Core i5 CPU and 16GB of memory.
We picked two families of graphs: balanced binary trees and 2D grids, and measured the convergence time as the size of the graph grows.
For balanced binary trees, we start with a tree with $4$ levels ($15$ nodes) and gradually increase the depth to $12$ levels ($4095$ nodes). 
For grids, we start with a grid of size $16 \times 2$ ($32$ nodes) and double the number of columns until we have a grid of size $16 \times 256$ ($4096$ nodes). 
For each criterion, we randomly initialize nodes in the layout from standard Gaussian, optimize the layout with respect to only one criterion using SGD and stop as soon as the layout converges. 
We ensure the convergence by gradually decreasing the learning rate of SGD: we decrease the learning rate by a factor of $0.7$ every time the loss has not decreased for a certain number of iterations, often referred to as the ``patience.'' 
In general, we need more patience for larger graphs and for smaller mini-batch to compensate for the large variance in loss estimation due to random sampling. 
Here, we set the patience to $max(100, int(|V|/m)*300)$ iterations when we optimize a graph with $|V|$ nodes and take $m$ samples in every SGD iteration.
\blue{To further improve the robustness of the stopping criteria, we smooth the sample loss by taking the exponential moving average.
That is, on the $i^{th}$ iteration, the smoothed loss $L_i$ is defined as
\begin{align*}
L_i &= \frac{SL_i + s^1 SL_{i-1} + \dots + s^{i-1} SL_1}{1+s+s^2+\dots+ s^{i-1}}
= \frac{\Sigma_{k=1}^i s^{i-k} \, SL_k}{\Sigma_{k=1}^i s^{i-k}}
\end{align*}
where $s$ is a smoothing factor. 
We set $s=0.5^{1/100}\approx 0.993$, a rate at which the $100^{th}$ preceding iteration will contribute half as much as that of the latest sample loss.
}


Fig.~\ref{fig:runtime} summarizes the runtime analysis for the two families of graphs (trees and grids) for all 9 criteria.  
Note that we are using log-log plots (log scales for both the $x$ and $y$ axes). This experimental analysis shows linear or near-linear time for the underlying algorithms. 
This is shown as steeper slopes in the log-log plots.

\section{Conclusions, Limitations, Future Work}

We introduced the graph drawing framework, \GDGD, for multicriteria graph drawing and
showed how this approach can be used to optimize different graph drawing criteria and combinations thereof. 
We showed that multiple readability criteria can be optimized jointly via SGD if each of them can be expressed as a differentiable function.
In cases that some readability criteria are not naturally differentiable (e.g., neighborhood preservation or crossing number), one can find differentiable surrogate functions and optimize the criteria indirectly. 
\blue{
We measured the quality of generated layouts 
and analyzed interactions between criteria, the runtime behavior, and the impact of sample sizes; all of which provide evidence of the effectiveness of \GDGD
}

\blue{\textbf{Support for More Constraint Types:}
Although \GDGD is a flexible framework that can optimize a wide range of criteria, we did not consider the class of constraints where the node coordinates are related by some inequalities (i.e., hard constraints). }
Similarly, in the \GDGD framework we do not naturally support shape-based drawing constraints such as those in~\cite{ipsepcola_2006, scalable_cola_2009,wang2017revisiting}. 
\blue{
Incorporating a wider range of constraint types and studying the interactions between them in the multi-objective setting are natural directions for future work.
}

\textbf{Better Weight Balancing for Multicriteria Objectives:}
The \GDGD framework is flexible and natural directions for future work include adding further drawing criteria and better ways to combine them.
\blue{
An appropriate balance between weights for the different criteria can be crucial as more and more criteria are incorporated into the optimization. 
Currently, we manually choose appropriate weight schedules based on specific combinations of criteria.
In the future, we would like to explore ways to automatically design and balance weight schedules in multicriteria graph drawing. 
}

\blue{\textbf{Applications of Different Techniques and Frameworks:}
Besides gradient descent, there are other optimization techniques
that could be deployed to multi-objective problems~\cite{orosz2020robust}. 
Similarly, while we used Tensorflow.js and PyTorch to implement \GDGD, there are other frameworks (e.g., pymoo~\cite{blank2020pymoo})  with support for multi-objective optimization. 
The application of different optimization techniques and  frameworks to multicriteria network visualization seems like an interesting direction for future work.
}

\blue{
\textbf{Scalability for Larger Graphs:}
Currently, not all criteria are fully optimized for speed.
Alternative objective functions, for example tsNET by Kruiger et al.~\cite{kruiger2017graph} for neighborhood preservation, could be considered in the \GDGD framework as further runtime scalability and quality improvements are needed for graphs with millions of nodes and edges.
One possible direction for improving scalability is to employ a multi-level algorithmic  framework.}

\section*{Acknowledgments}
This work was supported in part by NSF grants CCF-1740858, CCF-1712119, and
DMS-1839274.

\bibliographystyle{IEEEtranS}
\bibliography{references}

\vspace{-1cm}\begin{IEEEbiography}[{\vspace*{-.5cm}\includegraphics[width=.8in,height=1.0in,clip,keepaspectratio]{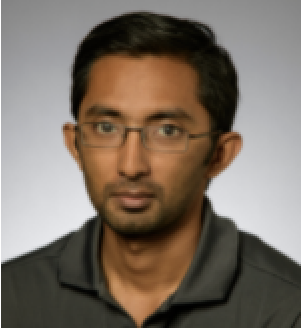}}]{Reyan Ahmed} is a Ph.D. student at the Department of Computer Science at the University of Arizona. He received B.Sc. and M.Sc. degree in Computer Science and Engineering from Bangladesh University of Engineering and Technology. His research interests include graph algorithms, network visualization, and data science.
\end{IEEEbiography}

\vspace{-1cm}\begin{IEEEbiography}[{\includegraphics[width=.8in,height=1.0in,clip,keepaspectratio]{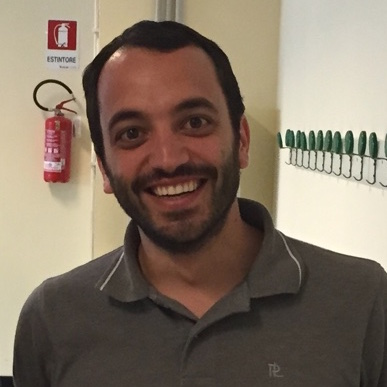}}]{Felice De Luca} is a postdoctoral researcher at the Department of Computer Science at the University of Arizona. He received an MS degree in 2014 and a PhD in 2018 at the Department of Computer and Automation Engineering at the University of Perugia. His research interests include graph drawing, information visualization, algorithm engineering, and computational geometry.
\end{IEEEbiography}

\vspace{-1cm}\begin{IEEEbiography}[{\vspace*{-1cm}\includegraphics[width=.8in,height=1.0in,clip,keepaspectratio]{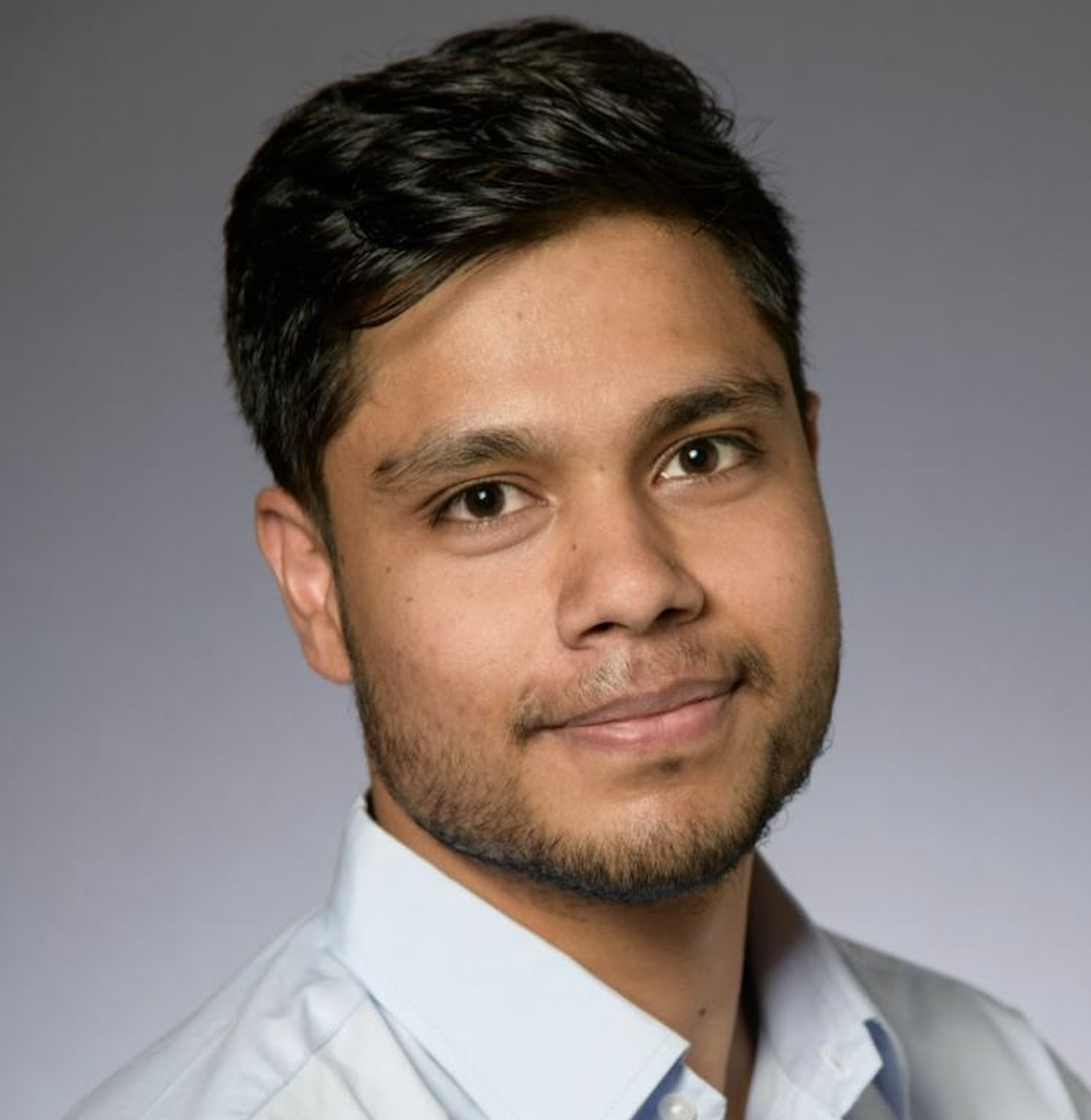}}]{Sabin Devkota} is a Ph.D. student at the Department of Computer Science at the University of Arizona. He received a bachelor's in Electronics and Communication Engineering from Tribhuvan University. His research interests include information visualization.
\end{IEEEbiography}

\vspace{-13cm}\begin{IEEEbiography}[{\vspace*{-.5cm}\includegraphics[width=.8in,height=1.0in,clip,keepaspectratio]{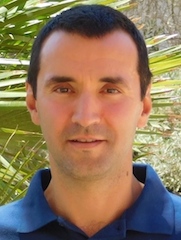}}]{Stephen Kobourov} is a Professor at the Department of Computer Science at the University of Arizona. He received a BS degree in Mathematics and Computer Science from Dartmouth College and MS and PhD degrees from Johns Hopkins University. His research interests include information visualisation, graph theory, and geometric algorithms.
\end{IEEEbiography}

\vspace{-13cm}\begin{IEEEbiography}[{\vspace*{-1cm}\includegraphics[width=.8in,height=1.0in,clip,keepaspectratio]{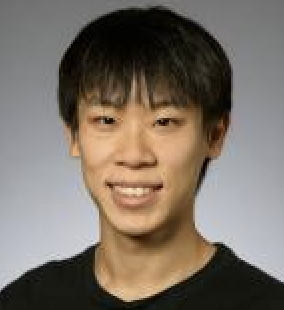}}]{Mingwei Li} is a PhD student in the Department of Computer Science, University of Arizona.
He received the BEng degree in electronics engineering from the Hong Kong University
of Science and Technology. 
His research interests include data visualization and machine learning.
\end{IEEEbiography}



\end{document}